\documentclass{aa}  

\usepackage{graphicx}
\usepackage{txfonts}
\usepackage{natbib}
\begin{document}

   \title{Near-infrared observations of RR Lyrae and Type II Cepheid variables in the metal-rich bulge globular cluster NGC 6441}


   \author{A. Bhardwaj\fnmsep\thanks{Marie Skłodowska-Curie Fellow}
          \inst{1}
         \and
          S. M. Kanbur\inst{2}
        \and
          M. Rejkuba\inst{3}
          \and
          M. Marconi\inst{1}
        \and
          M. Catelan\inst{4}
        \and
          V. Ripepi\inst{1}
         \and
          H. P. Singh\inst{5}
          }

   \institute{INAF-Osservatorio Astronomico di Capodimonte, Salita Moiariello 16, 80131, Naples, Italy\\
              \email{anupam.bhardwajj@gmail.com; anupam.bhardwaj@inaf.it}
         \and
        Department of Physics, State University of New York, Oswego, NY 13126, USA
         \and 
        European Southern Observatory, Karl-Schwarzschild-Stra\ss e 2, 85748, Garching, Germany
         \and
         Instituto de Astrof\`isica, Facultad de F\`isica, Pontificia Universidad Cat\`olica de Chile, Av. Vicu\~na Mackenna 4860, 7820436 Macul, Santiago, Chile
         \and
        Department of Physics and Astrophysics, University of Delhi, Delhi-110007, India
             }

   \date{Received August 10, 2022; accepted September 7, 2022}

 
  \abstract
   {NGC 6441 is a bulge globular cluster with an unusual horizontal branch morphology and a rich population of RR Lyrae (RRL) and Type II Cepheid (T2C) variables that is unexpected for its relatively high metallicity.}
   {Our goal is to characterize the pulsation properties of the population II RRL and T2C variables in a metal-rich globular cluster and compare them with a sample of variables in more metal-poor clusters as well as with theoretical predictions.  
   }
   {We present near-infrared (NIR, $JHK_s$) time-series observations of 42 RRL, 8 T2Cs, and 10 eclipsing binary candidate variables in NGC 6441. The multi-epoch observations were obtained using the FLAMINGOS-2 instrument on the 8-m Gemini South telescope. Multi-band data are used to investigate pulsation properties of RRL and T2C variables including their light curves, instability strip, period--amplitude diagrams, and period--luminosity and period--wesenheit relations (PLRs and PWRs) in $JHK_s$ filters. 
    }
   {NIR pulsation properties of RRL variables are well fitted with theoretical models that have canonical helium content and mean-metallicity of NGC 6441 ([Fe/H]$=-0.44\pm0.07$~dex). The helium-enhanced RRL models predict brighter NIR magnitudes and bluer colors than the observations of RRL in the cluster. We find that the empirical slopes of the RRL PLRs and PWRs in NGC 6441 are statistically consistent with those of RRL in more metal-poor globular clusters as well as the theoretical models. Therefore, we use theoretical calibrations of RRL period--luminosity--metallicity (PLZ) relations to simultaneously estimate the mean reddening, $E(J-K_s)=0.26\pm0.06$~mag, and the distance, $d=12.67\pm0.09$~kpc, to NGC 6441. Our mean reddening value is consistent with an independent estimate from the bulge reddening map based on red clump stars. The RRL based distance agrees well with similar literature measurements based on the PLZ relations, and the {\it Gaia} and other independent methods. Our distance and reddening values provide a very good agreement between the PLRs of T2Cs in NGC 6441 and those for RRL and T2Cs in Galactic globular clusters  that span a broad range of metallicity. 
   }
   {We conclude that NIR colour-magnitude diagram, pulsation properties, and PLRs for RRL and T2Cs in NGC 6441 are in good agreement with the predictions of RRL pulsation models with canonical helium content. This suggests that these population II variables are either not significantly helium enhanced as previously thought or the impact of such enhancement is smaller in NIR than the predictions of the pulsation models.}
   

\titlerunning{Variable stars in NGC 6441}
\authorrunning{Bhardwaj A. et al.}

   \maketitle
%

\section{Introduction}

Globular clusters (GCs) are important relics of the early formation history of the Galaxy. The Galactic bulge GCs are particularly interesting due to their diverse properties \citep{bica2016, minniti2017} and include a few peculiar systems. For example, Terzan 5 and Liller 1 host two distinct stellar populations with remarkably different ages similar to bulge fields \citep{ferraro2021} but unlike common Galactic GCs. This suggests that these strange GCs could be the fossil fragments from the hierarchical assembly of the Milky Way bulge \citep{ferraro2016}. 

A few peculiar metal-rich ($\rm{[Fe/H]}>-1.0$~dex) bulge GCs exhibit unique horizontal branch morphologies. Terzan 5, NGC 6569, and NGC 6440 display double horizontal branches \citep{mauro2012} while NGC 6441 and NGC 6388 have a very extended blue horizontal branch for their high metallicities \citep{rich1997}.
Unlike other metal-rich GCs which predominantly have a red horizontal branch, NGC 6441 and NGC 6388 host a significant population of RRL stars \citep{layden1999, pritzl2000, pritzl2003, skottfelt2015}. The red horizontal branch in NGC 6441 and NGC 6388 exhibits a tilt with the upward slope towards bluer colors in the optical color--magnitude diagrams \citep{rich1997,pritzl2003, catelan2006}. These clusters are suggested to harbor at least two stellar populations that are composed of stars with different helium abundances \citep{caloi2007, busso2007, bellini2013}. Despite many similarities between these twin clusters, the color--magnitude diagrams of NGC 6441 display a split main-sequences and no red horizontal branch structure in contrast to NGC 6388, indicating that the chemical patterns of different populations may be different in these two clusters \citep{bellini2013}.

NGC 6441 \citep[{[Fe/H]}$=-0.44\pm0.07$~dex,][]{carretta2009} is one of the most interesting metal-rich inner Galaxy GC due to its rich population of RRL variables \citep[81 RRL in the updated catalog of][]{clement2001}. The number of RRL in NGC 6441 is more than two times larger than that of known RRL in NGC 6388 \citep[37 RRL in][]{clement2001}. The RRL variables in these two clusters have exceptionally long periods that are not consistent with the typical pattern of decreasing mean-period with increasing mean-metallicities of Galactic GCs \citep{pritzl2000, catelan2009, bhardwaj2020}. The average fundamental-mode RRL period ($P_\textrm{RRab}$) in NGC 6441 and NGC 6388 is 0.76~d and 0.71~d, respectively. The $P_\textrm{RRab}$ for these metal-rich GCs is larger than that for RRL in Oosterhoff type I \citep[$P_\textrm{RRab} \sim$ 0.55~d and {[Fe/H]}$\gtrsim -1.5$~dex in OoI,][]{Oosterhoff1939,catelan2009, bhardwaj2022} and OoII GCs ($P_\textrm{RRab} \sim$0.65~d and {[Fe/H]}$\lesssim -1.5$~dex). These clusters have thus been considered the prototype of the third Oosterhoff type (OoIII) GCs \citep{pritzl2000}, and are peculiar cases of second-parameter effect at high metallicities \citep{catelan2009}. In addition to RRL, NGC 6441 and NGC 6388 also contain a relatively large number of T2Cs as compared to any other Milky Way GC regardless of their metallicities \citep{pritzl2003}. 

NGC 6441 is located in a very crowded bulge field and suffers not only from severe line of sight reddening \citep[$E(B-V)=0.47$~mag,][]{harris2010} but also from significant differential reddening \citep[$\Delta E(B-V)=0.14$~mag,][]{law2003}. While optical photometry of its variable stars is available \citep[e.g.,][]{layden1999, pritzl2001, pritzl2003, soszynski2014, skottfelt2015, soszynski2017}, the accurate determination of mean visual magnitude of RRL, and in turn the RRL-based distance estimate, is hampered by the uncertainties due to reddening and their peculiar evolutionary status. For example, \citet{pritzl2003} estimated distance to NGC 6441 between 10.4 kpc and 11.9 kpc assuming a range of absolute $V$-band magnitudes. \citet{harris2010} catalog lists a $V$-band distance modulus of 16.78 mag which results in a distance of 11.60 kpc to NGC 6441. NIR time-series photometry of RRL in NGC 6441 is limited to $K_s$-band data from \citet{dallora2008} and the Vista Variables in the Via Lactea (VVV) survey \citep{minnitivvv2010, alonsogarcia2021}. Using $K_s$-band photometry for RRL, both  \citet{dallora2008} and \citet{alonsogarcia2021} found a larger distance to NGC 6441 ($\gtrsim13\pm0.3$~kpc). Recently, \citet{baumgardt2021} combined astrometric and photometric data from {\it Gaia}, {\it Hubble Space Telescope (HST)} and literature to determine accurate distances to GCs. They found an average value of 12.73$\pm$0.16 kpc to NGC 6441 which is closer to the distance estimates based on NIR PLZ relations for RRL stars. NIR photometry is not only useful for constraining the reddening and the distance but also to investigate the impact of composition, in particular of possible helium enhancement, on RRL pulsation properties in NGC 6441.

In this paper, we present simultaneous NIR ($JHK_s$) multi-epoch photometry of RRL and T2Cs in NGC 6441 for the first time. The pulsation properties of RRL and T2C variables are discussed at NIR wavelengths in this metal-rich globular cluster complementing similar studies on metal-poor GCs \citep[M53 and M15 in][]{bhardwaj2021, bhardwaj2021a}. Section~\ref{sec:data} describes the NIR data, the data reduction, and the astrometric and photometric calibration. Section~\ref{sec:var} presents the variable stars and their NIR pulsation properties. The RRL and T2C PLRs as well as distance and reddening estimates for NGC 6441 are presented in Section~\ref{sec:plrs}. Section~\ref{sec:discuss} summarizes the main results of this work.\\

\section{Data and Photometry} \label{sec:data}

\subsection{Observations and data reduction}

Our NIR observations of NGC 6441 were obtained from 15 May 2021 to 08 July 2021 using the FLAMINGOS-2 instrument mounted on the 8-m Gemini South Telescope. FLAMINGOS-2 is a NIR imaging spectrograph with a $6.1\arcmin$ circular field of view and a pixel scale of $0.18\arcsec$ pixel$^{-1}$. We requested time-series observations in queue mode centered on the NGC 6441 cluster and obtained 21, 18, and 17 epochs in $J$, $H$, and $K_s$, respectively. Our observations for each epoch consisted of 5 dithered frames (exposures) on target and 5 dithered frames on an offset sky location in object-sky-object mode for all three filters. The offset sky exposures were necessary to ensure suitable background subtraction due to high density of stars in the object frames. The individual exposures were very short (5 sec) such that background sky variations between on- and off-target exposures were negligible.  Few epochs had more than 5 dithered frames, because some exposures were repeated due to changing conditions during the observations. A total of 307 (115 in $J$, 97 in $H$ and 95 in $K_s$) science frames are used in this study and the exposure time was 5 sec in all cases. A summary of observations is listed in Table ~\ref{tbl:data}.

\begin{table*}
\begin{center}
\caption{Log of NIR observations. \label{tbl:data}}
\scalebox{0.95}{
\begin{tabular}{cccccccccccc}
\hline\hline
{} &  &  \multicolumn{3}{c}{$J$-band}  &  \multicolumn{3}{c}{$H$-band}  &  \multicolumn{3}{c}{$K_s$-band}   & \\
{Date} & {MJD}	& {FWHM} & {AM} & {$N_\textrm{f}$} & {FWHM} & {AM} & {$N_\textrm{f}$} & {FWHM} & {AM} & {$N_\textrm{f}$} & ET \\
\hline
2021-05-15&   59349.301&     4.06&     1.07&  5&      ---&      ---& ---&      ---&      ---& ---&  5\\
2021-05-17&   59351.199&     3.86&     1.17&  5&     3.93&     1.15&  5&     3.80&     1.14&  5&  5\\
2021-05-17&   59351.398&     3.11&     1.02&  5&     3.31&     1.02&  5&     3.08&     1.02&  5&  5\\
2021-05-23&   59357.301&     5.11&     1.06&  5&      ---&      ---& ---&      ---&      ---& ---&  5\\
2021-05-23&   59357.398&     3.80&     1.07&  5&     3.89&     1.08&  5&     3.86&     1.09&  5&  5\\
2021-05-24&   59358.301&     3.79&     1.01&  5&     4.07&     1.01&  5&      ---&      ---& ---&  5\\
2021-05-24&   59358.398&     3.94&     1.17&  5&     3.96&     1.18&  5&     3.30&     1.20&  5&  5\\
2021-05-30&   59364.199&     3.90&     1.01&  5&     3.84&     1.01&  5&     3.75&     1.01&  5&  5\\
2021-05-31&   59365.199&     3.11&     1.01&  5&     3.28&     1.02&  5&     2.99&     1.02&  5&  5\\
2021-06-01&   59366.199&     3.62&     1.22&  5&     3.40&     1.19&  5&     4.08&     1.18&  5&  5\\
2021-06-01&   59366.301&     3.97&     1.04&  5&      ---&      ---& ---&      ---&      ---& ---&  5\\
2021-06-02&   59367.199&     3.47&     1.01&  5&     3.59&     1.01&  5&     3.30&     1.01&  5&  5\\
2021-06-02&   59367.301&     3.27&     1.08&  5&     3.61&     1.09&  5&     3.09&     1.10&  7&  5\\
2021-06-03&   59368.199&     3.89&     1.03& 10&     4.25&     1.02&  7&     3.93&     1.01&  8&  5\\
2021-06-12&   59377.000&     3.09&     1.60&  5&     3.31&     1.56&  5&     2.61&     1.52&  5&  5\\
2021-06-13&   59378.301&     4.35&     1.12&  5&     4.90&     1.13&  5&     3.86&     1.15&  5&  5\\
2021-06-14&   59379.000&     3.43&     1.54&  5&     3.75&     1.50&  5&     3.47&     1.47&  5&  5\\
2021-06-14&   59379.102&     3.79&     1.10&  5&     4.00&     1.09&  5&     3.62&     1.08&  5&  5\\
2021-06-14&   59379.199&     4.15&     1.04& 10&     4.20&     1.05& 10&     4.75&     1.07& 10&  5\\
2021-06-28&   59393.199&     3.95&     1.29&  5&     3.99&     1.32&  5&     3.69&     1.35&  5&  5\\
2021-07-08&   59403.102&     3.80&     1.10&  5&     3.77&     1.09&  5&     3.15&     1.08&  5&  5\\
\hline
\end{tabular}}
\end{center}
\footnotesize{{\bf Notes--} MJD: Modified Julian Date (JD$-$2,400,000.5). FWHM: Measured median full width at half maximum using {SExtractor}. AM: Median airmass. $N_\textrm{f}$: Number of dithered frames on target per epoch. ET: Exposure time (in seconds) for each dithered frame.}
\end{table*}

NIR science images and associated calibrations (dark and flat frames) were downloaded from the Gemini Observatory Archive{\footnote{\url{https://archive.gemini.edu/}}}. The DRAGONS (Data Reduction for Astronomy from Gemini Observatory North and South) pipeline was used for the pre-processing of images (bad-pixel mask, non-linearity correction, dark subtraction, and flat-fielding) and to obtain sky background images. However, we did not use DRAGONS to stack images because residual variations were present around bright stars in the derived sky background images. The residual variations were further removed using \texttt{SExtractor} \citep{bertin1996} before applying background subtraction.  The background corrected individual dithered frames were used for further analysis instead of the stacked mosaic at each epoch.

 We also created an astrometrically calibrated median-combined image from the $J$-band dithers obtained in the best-seeing epoch as a reference image. The $J$-band frames were selected because of an apparent saturation issue for the brightest stars in $H/K_s$ band in the best-seeing images. The individual dithered frames were astrometrically calibrated using an input source catalog generated with \texttt{SExtractor} together with the Two Micron All Sky Survey \citep[2MASS,][]{cutri2003} catalog in the \texttt{SCAMP} \citep{bertin2006}. Finally, a median-combined reference image was produced at the instrument pixel scale using \texttt{SWARP} \citep{bertin2006}. 
 
\subsection{Photometry}

The point-spread function (PSF) photometry  was performed on each image using \texttt{DAOPHOT/ALLSTAR} \citep{stetson1987} and \texttt{ALLFRAME} \citep{stetson1994} routines. In a given filter, all point sources with a brightness above $5\sigma$ of the detection threshold were found using \texttt{DAOPHOT}, and an aperture photometry was performed within a 3-pixel aperture. An empirical PSF was obtained for each image using 200 bright, isolated, and non-saturated stars excluding those located within central 200 pixels ($\sim0.6\arcmin$). 
The PSF photometry was obtained for the detected sources in all images using \texttt{ALLSTAR}. A common reference star-list was obtained from the astrometrically calibrated median-combined image for all three filters. Frame-to-frame coordinate transformations between the reference image and all epoch images in each NIR filters were derived using \texttt{DAOMATCH/DAOMASTER}. Finally, the reference-star list and the coordinate transformations were used as input for the PSF photometry to \texttt{ALLFRAME} performing profile fitting of all sources in the reference-star list across all the frames, simultaneously. We selected 50 secondary standards from the output photometry that are present in all frames in a given NIR filter. These secondary standards have small photometric uncertainties ($<0.005$~mag) and no epoch-to-epoch variability (rms $<0.01$~mag). Epoch dependent  frame-to-frame zero-point variations were corrected using these secondary standards to obtain catalogs with instrumental photometry in each NIR filter. 

\begin{figure}
  \includegraphics[width=\columnwidth]{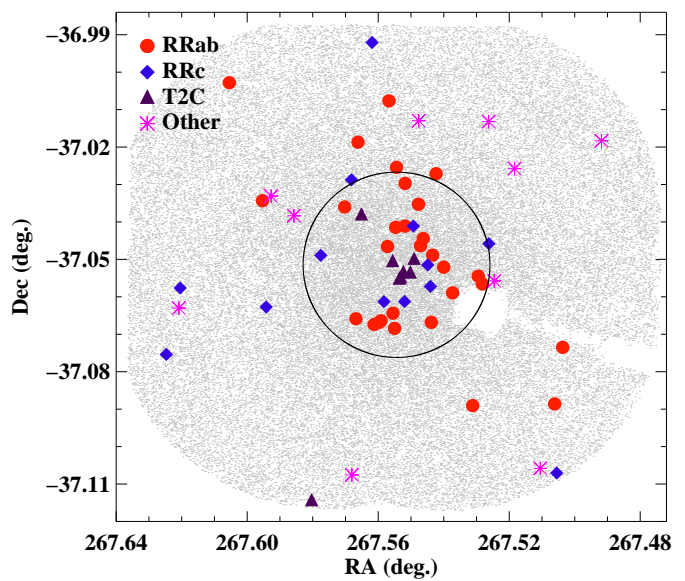}
  \caption{Spatial distribution of all point-like sources (grey dots) obtained from the reference mosaic $J$-band image covering a circular region of $\sim4.25'$ radius around the cluster center. The variable stars analysed in this work are shown with larger symbols according to the legend. The circle represents $3r_h$, where $r_h=0.57\arcmin$ is the half-light radius of the cluster \citep{harris2010}.}
  \label{fig:fov}
\end{figure}

\begin{figure*}
\includegraphics[width=\textwidth]{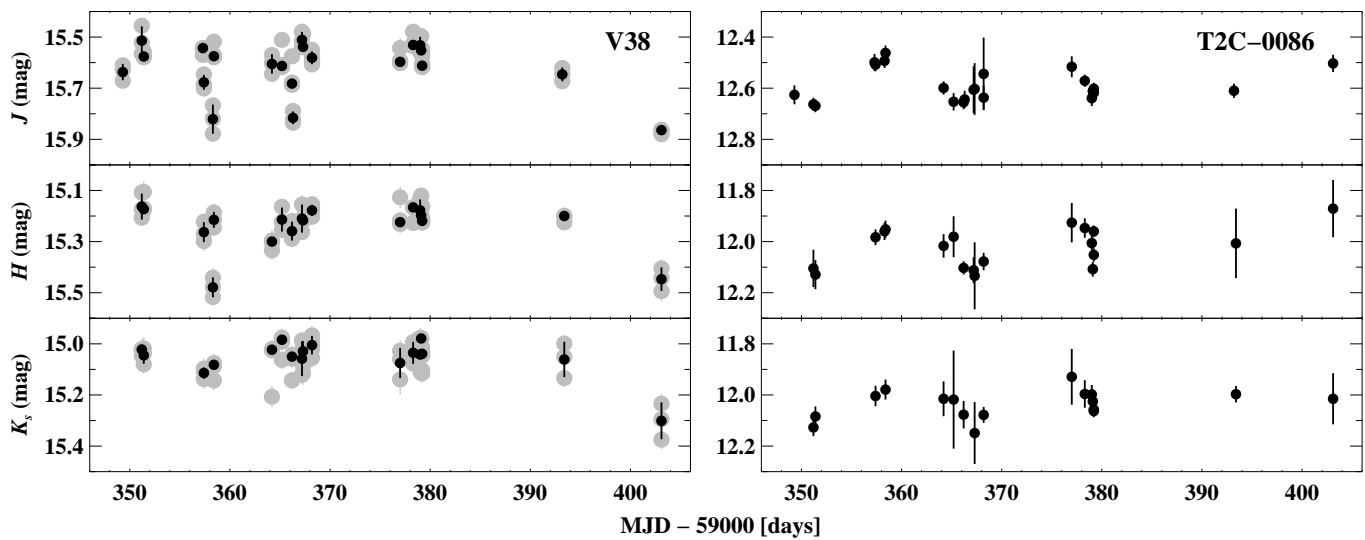}
\caption{NIR light curves of a RRL variable (left) found in all dithered frames  and a T2C (right) located in the outskirt of one of the dithered frames at each epoch. {\it Left:} Gray symbols show all photometric measurements (see $N_f$ columns in Table~\ref{tbl:data}) obtained from the dithered frames at a given epoch, and the black symbols represent the weighted mean magnitudes. {\it Right:} No weighted averaging was performed.} 
\label{fig:bin_lcs}
\end{figure*}

The photometric calibration was performed using 2MASS catalog in $JHK_s$ bands. Our NIR photometry was cross-matched with the 2MASS within a tolerance of $1.0\arcsec$. We found 1783 stars in common between 2MASS and our NIR catalogs. Due to the crowded nature of the cluster, the sample was restricted to stars with 2MASS quality flag ``A'' and having photometric uncertainties $<0.05$~mag. Furthermore, only stars within magnitude ranges $12.0<J<16.0$~mag and $11.5<K_s<15.5$~mag were considered to avoid saturation and non-linearity at the bright end. Finally, a clean sample of 390 stars was used for photometric calibration after excluding sources located within the central $1\arcmin$ of the cluster. We solved for a magnitude zero-point and a color term for the calibration but adding the linear color term did not contribute to any reduction in the root mean square (rms) error or the chi-squared per degree of freedom. Since the color coefficients were found to be consistent with zero within their uncertainties, the absolute calibration was obtained by correcting only for the fixed zero-point offset between instrumental and 2MASS magnitudes. A comparison with the VVV data \citep[][private comm.]{alonsogarcia2021} results in a difference of $\Delta J=0.022$~mag and $\Delta K_s=0.043$~mag between VISTA and our calibrated photometry for the aforementioned magnitude ranges. This difference increases to $\Delta J=0.046$~mag and $\Delta K_s=0.050$~mag if the VVV photometry is transformed to 2MASS using the equations provided in \citet{fernandes2018}.

\section{Variable stars in NGC 6441}
\label{sec:var}

The variable star population of NGC 6441 has been presented in detail in optical studies by \citet{layden1999}, \citet{pritzl2001, pritzl2003}, and \citet{skottfelt2015}. Recently, $ZYJK_s$ photometry of variable stars in NGC 6441 was provided by \citet{alonsogarcia2021} which includes well-sampled light curves only in $K_s$-band from the VVV survey.  We compiled a list of RRL, T2C, and eclipsing binary variables with periods less than 50 days from the updated catalog of \citet{clement2001}{\footnote{\url{http://www.astro.utoronto.ca/~cclement/}}}, Optical Gravitational Lensing Experiment survey \citep[OGLE,][]{soszynski2014, soszynski2017}, and \citet{alonsogarcia2021}. Within our mosaic field of view, there are 89, 68, and 33 variables from the aforementioned three studies. \citet{clement2001} catalog has 48 and 22 common variables with OGLE and \citet{alonsogarcia2021}, respectively. After excluding common variables among three sources, our initial sample was of 112 variables including 77 RRL and 9 T2Cs. Only two variables (V127 and V151) among these are in close vicinity of $1\arcsec$ radius. The periods and optical $V/I$ band magnitudes and amplitudes were also adopted, if available. 

We extracted $JHK_s$ light curves for each variable in our sample by cross-matching with our catalog within an initial tolerance of $1\arcsec$. 
The light curves in each NIR filter were phased with known pulsation periods and were visually inspected to confirm their variability and periodicity.
The initial cross-matching radius was increased to $2\arcsec$ if no variable source was found. We found 60 candidate variables to exhibit clear periodic variations and evidence of variability. Out of remaining 52 candidates, 30 variables fall within the crowded unresolved central $0.5\arcmin$ region where those were discovered in {\it HST} images \citep{pritzl2003}. Among the variables located outside the crowded region, we only missed 6 candidate RRL (V40, V70, V102, V123, V124, and V149). Note that V40, V102, and V149 were also not recovered in OGLE survey and are unlikely variable. While V70 falls on the guide probe in our images (see Fig.~\ref{fig:fov}) which compromises its photometry, V123 and V124 are both blended with a bright source. Most of the known T2C (8 out of 9) light curves were retrieved except for the only short-period ($P=2.5474$ days) BL Herculis variable in NGC 6441. For a comparison, \citet{alonsogarcia2021} recovered only 2 RRL within central $1\arcmin$ region of the cluster while we retrieved light curves of 17 known RRL variables thanks to the better spatial resolution of FLAMINGOS-2. Since our cadence and exposure times were primarily aimed at bright targets, many fainter eclipsing binaries were not recovered. However, we obtained NIR light curves of 5 OGLE eclipsing binary and 5 unclassified candidate variables from \citet{alonsogarcia2021}, which are included in the candidate eclipsing binary sample in this work. Our final sample of 60 variables in NGC 6441 includes 42 RRL, 8 T2C, and 10 eclipsing binary variables. Fig.~\ref{fig:fov} shows the spatial distribution of these variables on the reference image. 

\subsection{Light curves and templates}

\begin{figure*}
\centering
\includegraphics[width=0.92\textwidth]{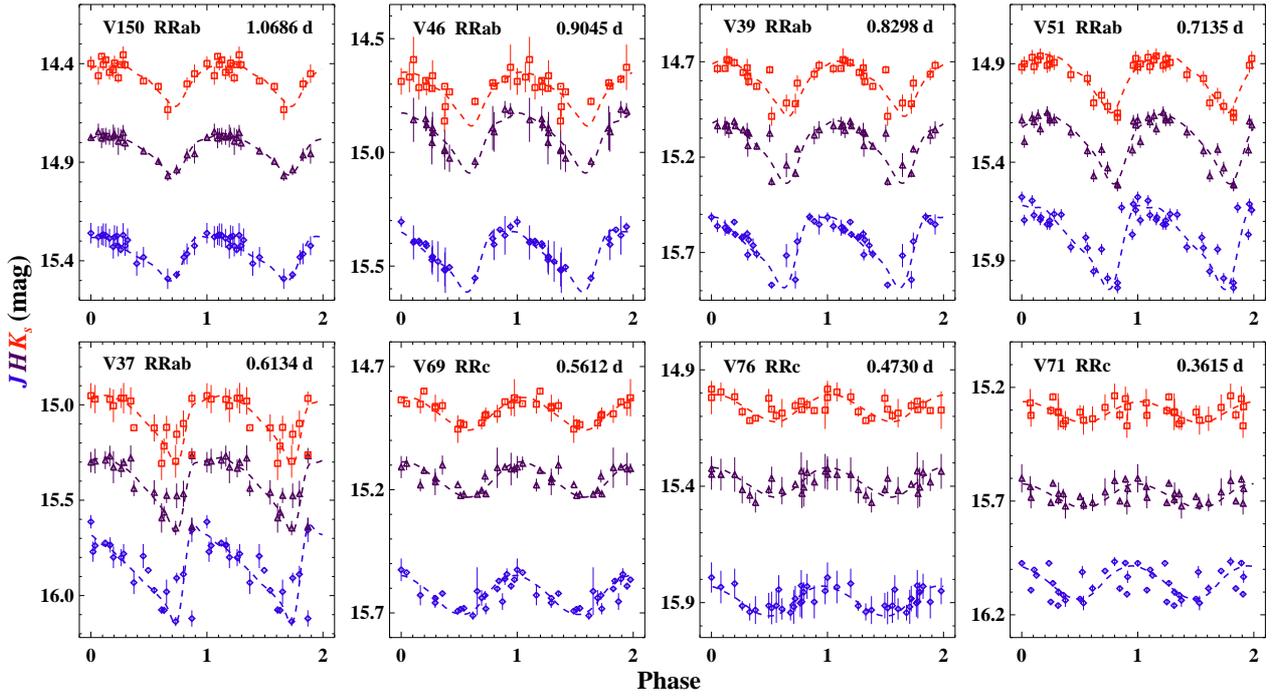}
\caption{Example phased light curves of RRL stars in NIR bands covering the entire range of periods in our sample. The $J$ (blue) and $K_s$-band (red)  light curves are offset for clarity by $+0.1$ mag and $-0.2$~mag, respectively. The dashed lines represent the best-fitting templates to the data in each band. Star ID, variable subtype, and the pulsation period are included at the top of each panel.} 
\label{fig:lcs_rrl}
\end{figure*}

\begin{figure*}
\centering
\includegraphics[width=0.92\textwidth]{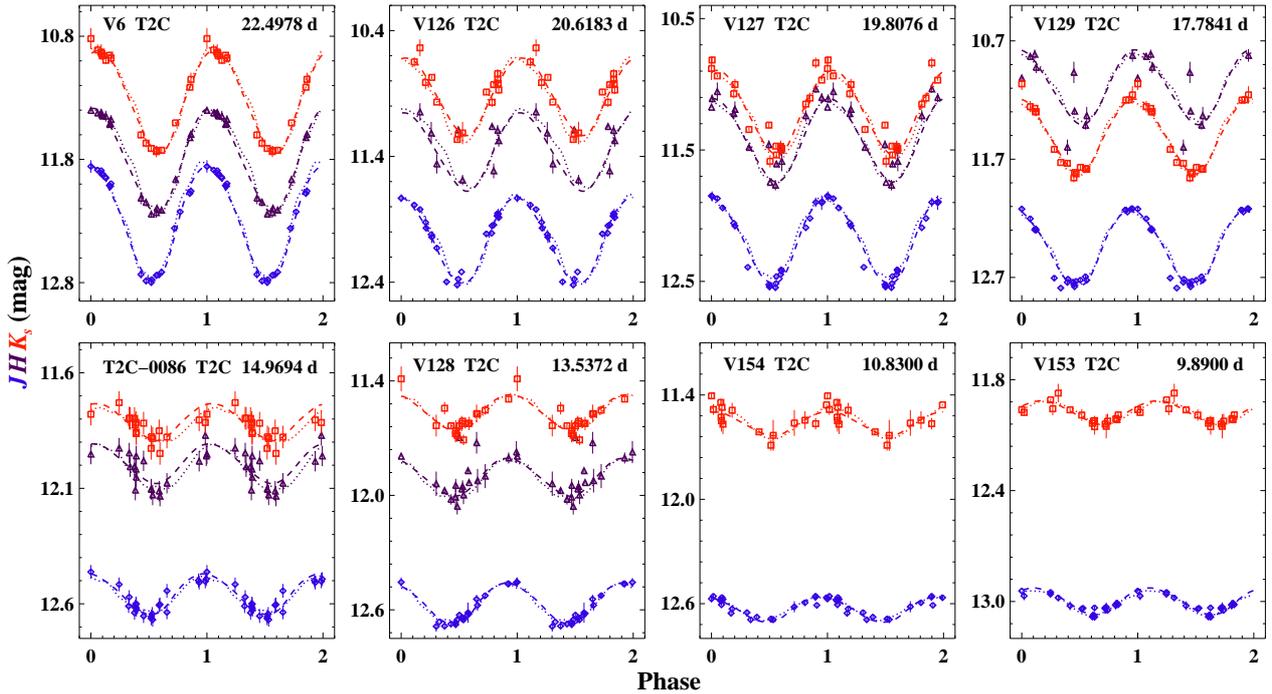}
\caption{Phased light curves of all T2C candidates in our sample. Only the $K_s$-band (red) light curves are offset for clarity by $-0.2$~mag. The dashed and dotted lines represent the best-fitting sinusoidal template and $K_s$-band T2C templates \citep{bhardwaj2017} to the data. Star ID, variable subtype, and the pulsation period are included at the top of each panel.} 
\label{fig:lcs_cep}
\end{figure*}

\begin{figure*}
\centering
\includegraphics[width=0.95\textwidth]{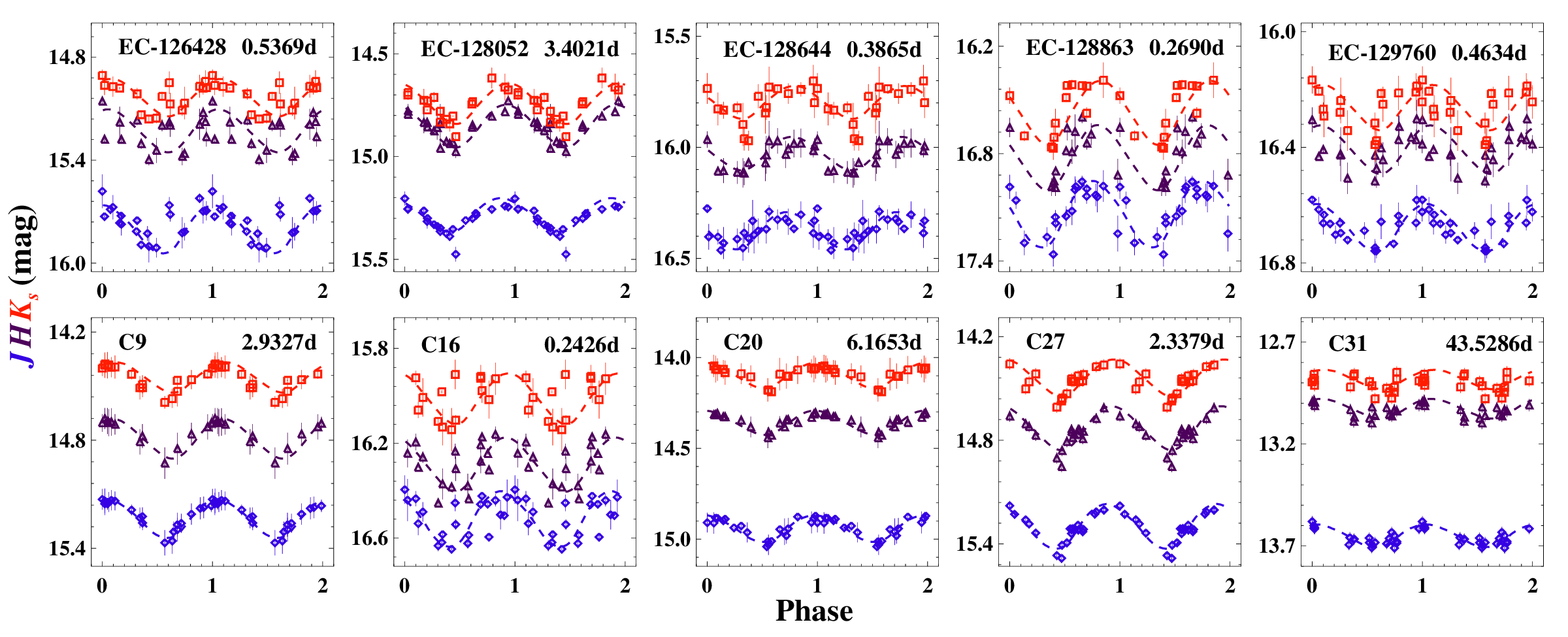}
\caption{Phased light curves of five eclipsing binary and five unclassified variables in $JHK_s$-bands. The dashed lines represent the best-fitting sinusoidal templates to the data. Star ID, variable subtype, and the pulsation period are included at the top of each panel.} 
\label{fig:lcs_ec}
\end{figure*}

We performed photometry on each dithered frame and most of the candidate variables within FLAMINGOS-2 field of view were observed in all 307 $JHK_s$ science frames. As the total integration time per epoch is significantly smaller than the variability timescale of RRL and T2C, we take a weighted mean of magnitudes obtained from all frames for a given filter at a given epoch. The standard deviation of the mean was propagated to the photometric uncertainties. However, a few most outskirt variables were found only in 1-2 dithered frames per epoch and no weighted averaging was performed in these cases. 
Figure~\ref{fig:bin_lcs} shows example binned and non-binned light curves of a central and an outskirt variable, respectively. Table~\ref{tbl:phot_lcs} provides the time-series photometry for all variables analysed in this work.

\begin{table}
\begin{center}
\caption{Time-series photometry of variables in NGC 6441. \label{tbl:phot_lcs}}
\begin{tabular}{ccccc}
\hline\hline
{ID} & {Band} & {MJD} & {Mag.} & {$\sigma_{\textrm{mag}}$}\\
\hline
           V6&    $J$&    59377.000&   11.857&    0.051\\
           V6&    $J$&    59378.301&   11.886&    0.003\\
           V6&    $J$&    59379.000&   11.890&    0.006\\
     ... &   ...&         ... &      ... &      ... \\
           V6&    $H$&    59377.000&   11.402&    0.011\\
           V6&    $H$&    59378.301&   11.424&    0.025\\
           V6&    $H$&    59379.000&   11.420&    0.028\\
     ... &   ...&         ... &      ... &      ... \\
           V6&  $K_s$&    59377.000&   11.019&    0.085\\
           V6&  $K_s$&    59378.301&   11.111&    0.044\\
           V6&  $K_s$&    59379.000&   11.129&    0.064\\
     ... &   ...&         ... &      ... &      ... \\
\hline
\end{tabular}
\end{center}
\footnotesize{{\bf Notes- }ID: IDs in the catalog of \citet{clement2001} or \citet{soszynski2014} or \citet{alonsogarcia2021}; MJD $=$ JD $- 2,400,000.5$. The fourth and fifth columns represent magnitude and its associated uncertainty in the given band. This table is available in its entirety in machine-readable form.}
\end{table}

NIR light curves of RRL variables were fitted with $JHK_s$ templates from \citet{braga2019}. The available templates were constructed using NIR light curves of RRL in $\omega$ Cen which cover three period bins for RRab and a single period bin for all RRc stars.  Since RRL in NGC 6441 have unusually long periods, we fitted all available NIR templates to RRL where period-bin based template did not fit all the data points in the light curves. In these cases, the best-fitting templates were obtained based on chi-square minimization. For T2Cs, we fitted two sets of templates - a sinusoidal template and $K_s$-band templates from \citet{bhardwaj2017a}. The $K_s$ band templates were rescaled to fit the $J$ and $H$ band light curve by solving for the amplitudes, phase offsets, and mean-magnitudes, simultaneously. The difference in the mean magnitudes obtained from the two sets of templates is significantly smaller than their photometric uncertainties. For candidate eclipsing binaries, sinusoidal templates were used to fit NIR light curves.

\begin{figure}
  \includegraphics[width=\columnwidth]{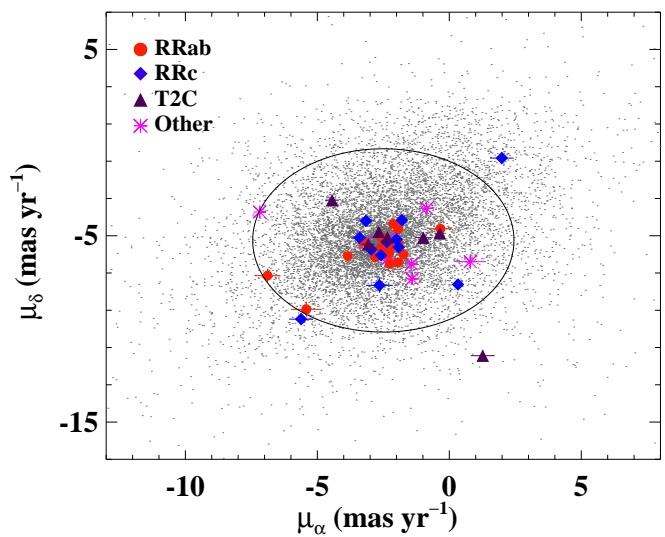}
  \caption{Proper motions of variable stars in NGC 6441 from {\it Gaia} DR3. The ellipse represents three times the width of the Gaussian distributions along both the proper motion axis. The outlier RRL (V60) and T2C (V154) have unreliable astrometric solution (ruwe$~>5$).}
  \label{fig:pms}
\end{figure}

Figs.~\ref{fig:lcs_rrl}, \ref{fig:lcs_cep}, and \ref{fig:lcs_ec} display NIR light curves and fitted templates to representative RRL stars with different periods, all T2C variables, and eclipsing binary candidates, respectively. The unclassified variables (C9, C16, C20, C27, and C31) from \citet{alonsogarcia2021} are shown with candidate eclipsing binary variables. A larger scatter is evident in the light curves of fainter RRc and eclipsing binary candidates. Some of the bright T2Cs are in the non-linearity limit of the FLAMINGOS-2, in particular in the $H$-band. The light curve of V129 in $H$-band is brighter than $K_s$ while no variability was seen for V153 and V154 in $H$-band. These two centrally located T2Cs are likely saturated. The $H$-band photometry of these three T2Cs will not be used further in this analysis. The best-fitted templates were used to determine intensity averaged mean magnitudes and peak to peak amplitudes of all variables. Median photometric and the rms uncertainties of the templates were added to the errors on the mean magnitudes and amplitudes obtained from the best-fitting templates, respectively. Table~\ref{tbl:var} lists the pulsation properties of variable stars analysed in this work.

\begin{sidewaystable*}
\begin{center}
\caption{NIR pulsation properties of variable stars in NGC 6441. \label{tbl:var}}
\scalebox{0.8}{
\begin{tabular}{lccclccccccccccccccc}
\hline\hline
{ID} & {RA} & {Dec} & {$P$} &  {Type}& \multicolumn{3}{c}{Mean magnitudes ($m_\lambda$)}  & \multicolumn{3}{c}{$\sigma_{m_\lambda}$} & \multicolumn{3}{c}{Amplitudes ($Amp_\lambda$) }  & \multicolumn{3}{c}{$\sigma_{Amp_\lambda}$}& {$E_{JK_s}$} &{$\sigma_{E_{JK_s}}$} & {OTH ID}\\
 	&	&   &    &	   & $J$  &   $H$  & $K_s$  & $J$  &  $H$  & $K_s$  &  $J$  &   $H$  & $K_s$   & $J$  &   $H$  & $K_s$	& & &\\  
 	&	deg.&	deg.	& days 	    &	   & \multicolumn{3}{c}{mag}  & \multicolumn{3}{c}{mag}  & \multicolumn{3}{c}{mag}   &  \multicolumn{3}{c}{mag}  & mag &  mag &  \\
\hline
         V6&    267.56513&    -37.03783&   22.49778&        T2C&   12.253&   11.760&   11.473&    0.054&    0.051&    0.055&    0.923&    0.845&    0.833&    0.060&    0.041&    0.049&    0.186&    0.025& T2C-0083\\
         V37&    267.56617&    -37.01869&    0.61338&       RRab&   15.746&   15.400&   15.263&    0.028&    0.027&    0.027&    0.514&    0.390&    0.362&    0.122&    0.091&    0.133&    0.205&    0.019& RRL-03991\\
         V38&    267.55671&    -37.00767&    0.73460&       RRab&   15.651&   15.265&   15.094&    0.025&    0.023&    0.023&    0.357&    0.320&    0.311&    0.105&    0.078&    0.091&    0.208&    0.020& RRL-03967\\
         V39&    267.59533&    -37.03433&    0.82979&       RRab&   15.548&   15.129&   15.002&    0.025&    0.024&    0.024&    0.378&    0.329&    0.289&    0.058&    0.058&    0.056&    0.295&    0.019& RRL-04057\\
         V41&    267.56679&    -37.06586&    0.74760&       RRab&   15.698&   15.313&   15.156&    0.025&    0.024&    0.024&    0.387&    0.327&    0.311&    0.061&    0.055&    0.083&    0.255&    0.021& RRL-03994\\
         V42&    267.55442&    -37.02542&    0.81264&       RRab&   15.537&   15.145&   14.979&    0.024&    0.024&    0.023&    0.291&    0.282&    0.217&    0.036&    0.067&    0.056&    0.160&    0.017& RRL-03956\\
         V43&    267.57025&    -37.03600&    0.77307&       RRab&   15.580&   15.177&   14.989&    0.023&    0.023&    0.022&    0.264&    0.220&    0.219&    0.048&    0.037&    0.055&    0.202&    0.019& RRL-04003\\
         V44&    267.55917&    -37.06639&    0.60546&       RRab&   14.381&   13.746&   13.509&    0.020&    0.020&    0.021&    0.117&    0.080&    0.121&    0.023&    0.043&    0.053&    0.242&    0.015& ---\\
         V45&    267.53117&    -37.08906&    0.50338&       RRab&   15.415&   15.038&   15.042&    0.027&    0.023&    0.023&    0.334&    0.175&    0.190&    0.072&    0.049&    0.067&    0.296&    0.023& RRL-03886\\
         V46&    267.50367&    -37.07350&    0.90449&       RRab&   15.345&   14.925&   14.934&    0.024&    0.026&    0.024&    0.272&    0.264&    0.237&    0.027&    0.028&    0.055&    0.246&    0.016& RRL-03813\\
         V51&    267.60546&    -37.00281&    0.71354&       RRab&   15.678&   15.284&   15.161&    0.035&    0.035&    0.031&    0.425&    0.356&    0.296&    0.080&    0.050&    0.035&    0.256&    0.014& RRL-04080\\
         V52&    267.54233&    -37.02711&    0.85909&       RRab&   15.508&   15.108&   14.935&    0.022&    0.022&    0.022&    0.162&    0.155&    0.151&    0.030&    0.029&    0.038&    0.257&    0.021& RRL-03916\\
         V53&    267.55179&    -37.02972&    0.85205&       RRab&   15.462&   15.041&   14.838&    0.022&    0.022&    0.021&    0.154&    0.148&    0.130&    0.027&    0.050&    0.022&    0.184&    0.022& RRL-03945\\
         V55&    267.55183&    -37.04111&    0.69749&       RRab&   15.538&   15.138&   14.952&    0.027&    0.025&    0.028&    0.417&    0.316&    0.298&    0.088&    0.068&    0.070&    0.187&    0.024& RRL-03946\\
         V56&    267.54625&    -37.04442&    0.90561&       RRab&   14.166&   13.506&   13.218&    0.020&    0.020&    0.020&    0.116&    0.097&    0.099&    0.023&    0.036&    0.029&    0.189&    0.027& ---\\
         V57&    267.54338&    -37.04886&    0.69600&       RRab&   15.590&   15.180&   14.981&    0.027&    0.024&    0.024&    0.458&    0.344&    0.339&    0.059&    0.073&    0.056&    0.239&    0.021& RRL-03918\\
         V59&    267.53729&    -37.05894&    0.70285&       RRab&   15.694&   15.310&   15.208&    0.027&    0.024&    0.025&    0.423&    0.347&    0.339&    0.139&    0.077&    0.063&    0.228&    0.023& RRL-03904\\
         V60&    267.55550&    -37.06439&    0.85350&       RRab&   15.368&   14.679&   14.445&    0.020&    0.020&    0.021&    0.075&    0.056&    0.090&    0.018&    0.013&    0.023&    0.224&    0.023& RRL-03964\\
         V61&    267.56125&    -37.06739&    0.75011&       RRab&   15.641&   15.184&   15.032&    0.024&    0.024&    0.023&    0.301&    0.287&    0.257&    0.120&    0.054&    0.063&    0.232&    0.016& RRL-03980\\
         V62&    267.55500&    -37.06844&    0.67997&       RRab&   15.706&   15.340&   15.218&    0.030&    0.030&    0.028&    0.460&    0.388&    0.344&    0.116&    0.077&    0.096&    0.257&    0.015& RRL-03961\\
         V63&    267.54708&    -37.04636&    0.69781&       RRab&   15.180&   14.680&   14.437&    0.022&    0.022&    0.022&    0.225&    0.222&    0.226&    0.059&    0.050&    0.068&    0.239&    0.025& RRL-03933\\
         V64&    267.54000&    -37.05208&    0.71710&       RRab&   15.485&   15.118&   14.866&    0.026&    0.026&    0.023&    0.353&    0.347&    0.253&    0.078&    0.111&    0.085&    0.146&    0.024& RRL-03912\\
         V65&    267.55467&    -37.04144&    0.75850&       RRab&   15.375&   14.960&   14.719&    0.022&    0.022&    0.022&    0.203&    0.184&    0.206&    0.063&    0.035&    0.036&    0.216&    0.022& RRL-03957\\
         V66&    267.54771&    -37.03528&    0.86090&       RRab&   15.492&   15.080&   14.877&    0.022&    0.022&    0.022&    0.200&    0.187&    0.165&    0.032&    0.055&    0.038&    0.219&    0.023& RRL-03934\\
         V68&    267.50550&    -37.10706&    0.32373&        RRc&   14.039&   13.642&   13.760&    0.024&    0.021&    0.022&    0.149&    0.117&    0.112&    0.054&    0.046&    0.030&    0.252&    0.014& RRL-03816\\
         V69&    267.62046&    -37.05764&    0.56119&        RRc&   15.523&   15.156&   15.088&    0.023&    0.023&    0.021&    0.158&    0.143&    0.132&    0.034&    0.034&    0.023&    0.304&    0.014& RRL-04118\\
         V71&    267.56817&    -37.02875&    0.36155&        RRc&   15.954&   15.676&   15.507&    0.022&    0.021&    0.021&    0.145&    0.107&    0.094&    0.062&    0.040&    0.045&    0.217&    0.020& RRL-03996\\
         V72&    267.56187&    -36.99208&    0.31167&        RRc&   16.034&   15.805&   15.718&    0.026&    0.023&    0.022&    0.193&    0.147&    0.122&    0.075&    0.044&    0.068&    0.251&    0.024& RRL-03982\\
         V74&    267.57754&    -37.04892&    0.31761&        RRc&   15.941&   15.588&   15.413&    0.022&    0.022&    0.021&    0.151&    0.133&    0.109&    0.075&    0.039&    0.087&    0.295&    0.025& RRL-04015\\
         V75&    267.55196&    -37.06125&    0.40500&        RRc&   15.880&   15.590&   15.457&    0.021&    0.020&    0.021&    0.113&    0.089&    0.089&    0.035&    0.036&    0.046&    0.245&    0.020& ---\\
         V76&    267.59421&    -37.06272&    0.47297&        RRc&   15.790&   15.381&   15.262&    0.021&    0.021&    0.021&    0.128&    0.128&    0.116&    0.039&    0.037&    0.040&    0.290&    0.027& RRL-04055\\
         V77&    267.62471&    -37.07539&    0.37410&        RRc&   15.830&   15.513&   15.476&    0.027&    0.023&    0.026&    0.178&    0.150&    0.148&    0.036&    0.039&    0.064&    0.246&    0.021& RRL-04137\\
         V79&    267.55825&    -37.06131&    0.41721&        RRc&   15.774&   15.420&   15.257&    0.022&    0.023&    0.023&    0.161&    0.144&    0.136&    0.040&    0.063&    0.071&    0.237&    0.024& RRL-03971\\
         V93&    267.52621&    -37.04581&    0.33825&        RRc&   15.994&   15.717&   15.569&    0.023&    0.021&    0.021&    0.206&    0.121&    0.092&    0.065&    0.048&    0.032&    0.179&    0.018& RRL-03870\\
         V94&    267.54929&    -37.04111&    0.38505&        RRc&   15.431&   15.001&   14.724&    0.021&    0.021&    0.021&    0.137&    0.121&    0.101&    0.072&    0.041&    0.042&    0.186&    0.030& RRL-03940\\
         V96&    267.50612&    -37.08861&    0.85448&       RRab&   15.562&   15.148&   14.999&    0.022&    0.023&    0.024&    0.161&    0.124&    0.132&    0.029&    0.031&    0.037&    0.255&    0.017& RRL-30154\\
         V97&    267.52821&    -37.05658&    0.84400&       RRab&   15.471&   15.042&   14.922&    0.021&    0.021&    0.022&    0.132&    0.132&    0.151&    0.028&    0.028&    0.043&    0.218&    0.028& ---\\
        V106&    267.54404&    -37.05722&    0.36109&        RRc&   15.440&   15.050&   14.843&    0.021&    0.022&    0.022&    0.184&    0.141&    0.136&    0.050&    0.051&    0.065&    0.202&    0.023& RRL-03921\\
        V126&    267.55233&    -37.05325&   20.61828&        T2C&   12.029&   11.337&   11.090&    0.035&    0.039&    0.034&    0.681&    0.623&    0.616&    0.056&    0.267&    0.074&    0.265&    0.022& T2C-1249\\
        V127&    267.55021&    -37.05336&   19.80759&        T2C&   12.155&   11.398&   11.377&    0.044&    0.041&    0.045&    0.650&    0.634&    0.641&    0.053&    0.178&    0.089&    0.265&    0.022& T2C-0752\\
        V128&    267.54904&    -37.04969&   13.53716&        T2C&   12.561&   11.901&   11.760&    0.022&    0.023&    0.022&    0.216&    0.198&    0.188&    0.020&    0.098&    0.045&    0.223&    0.033& T2C-1248\\
        V129&    267.55354&    -37.05500&   17.78405&        T2C&   12.415&   ---&   11.671&    0.045&    ---&    0.047&    0.650&    ---&    0.633&    0.047&    0.186&    0.047&    0.291&    0.024& T2C-1250\\
        V140&    267.54487&    -37.05150&    0.61418&        RRc&   15.355&   15.021&   14.801&    0.021&    0.021&    0.021&    0.152&    0.123&    0.120&    0.033&    0.031&    0.041&    0.164&    0.020& RRL-03923\\
        V143&    267.55717&    -37.04656&    0.86271&       RRab&   15.193&   14.877&   14.574&    0.022&    0.028&    0.024&    0.239&    0.256&    0.262&    0.065&    0.174&    0.066&    0.244&    0.031& RRL-03969\\
        V150&    267.52946&    -37.05447&    1.06862&       RRab&   15.262&   14.833&   14.675&    0.022&    0.022&    0.022&    0.226&    0.211&    0.224&    0.029&    0.019&    0.037&    0.202&    0.027& C8\\
        V153&    267.55321&    -37.05461&    9.89000&        T2C&   12.995&   ---&   12.174&    0.021&    ---&    0.021&    0.144&    ---&    0.127&    0.016&    0.123&    0.028&    0.291&    0.024& ---\\
        V154&    267.55558&    -37.05031&   10.83000&        T2C&   12.629&   ---&   11.761&    0.022&   ---&    0.023&    0.146&    ---&    0.137&    0.019&    0.221&    0.056&    0.312&    0.037& ---\\
   RRL-03920&    267.54379&    -37.06681&    0.64801&       RRab&   15.724&   15.338&   15.174&    0.026&    0.025&    0.024&    0.427&    0.352&    0.305&    0.068&    0.072&    0.058&    0.279&    0.023& ---\\
   RRL-03973&    267.55971&    -37.06681&    0.60880&       RRab&   14.379&   13.734&   13.528&    0.020&    0.020&    0.021&    0.133&    0.107&    0.106&    0.025&    0.045&    0.035&    0.242&    0.015& ---\\
    T2C-0086&    267.58033&    -37.11417&   14.96944&        T2C&   12.553&   11.990&   12.008&    0.024&    0.029&    0.024&    0.176&    0.171&    0.160&    0.026&    0.040&    0.031&    0.201&    0.027& ---\\
\hline
\end{tabular}}
\end{center}
\footnotesize{}
\end{sidewaystable*}

\begin{sidewaystable*}
\begin{center}
\scalebox{0.8}{
\begin{tabular}{lccclccccccccccccccc}
\hline\hline
{ID} & {RA} & {Dec} & {$P$} &  {Type}& \multicolumn{3}{c}{Mean magnitudes ($m_\lambda$)}  & \multicolumn{3}{c}{$\sigma_{m_\lambda}$} & \multicolumn{3}{c}{Amplitudes ($Amp_\lambda$) }  & \multicolumn{3}{c}{$\sigma_{Amp_\lambda}$}& {$E_{JK_s}$} &{$\sigma_{E_{JK_s}}$} & {OTH ID}\\
 	&	&   &    &	   & $J$  &   $H$  & $K_s$  & $J$  &  $H$  & $K_s$  &  $J$  &   $H$  & $K_s$   & $J$  &   $H$  & $K_s$	& & &\\  
 	&	deg.&	deg.	& days 	    &	   & \multicolumn{3}{c}{mag}  & \multicolumn{3}{c}{mag}  & \multicolumn{3}{c}{mag}   &  \multicolumn{3}{c}{mag}  & mag &  mag &  \\
\hline
   EC-126428&    267.51837&    -37.02575&    0.53691&         EC&   15.894&   15.223&   15.128&    0.024&    0.025&    0.023&    0.280&    0.250&    0.217&    0.087&    0.113&    0.061&    0.187&    0.024& ---\\
   EC-128052&    267.56804&    -37.10756&    3.40209&         EC&   15.379&   14.848&   14.836&    0.023&    0.024&    0.024&    0.163&    0.207&    0.196&    0.047&    0.041&    0.052&    0.254&    0.018& ---\\
   EC-128644&    267.58575&    -37.03836&    0.38646&         EC&   16.471&   16.025&   15.893&    0.022&    0.023&    0.024&    0.167&    0.150&    0.148&    0.040&    0.033&    0.056&    0.240&    0.026& ---\\
   EC-128863&    267.59271&    -37.03311&    0.26902&         EC&   17.224&   16.812&   16.657&    0.028&    0.030&    0.030&    0.371&    0.364&    0.358&    0.074&    0.062&    0.098&    0.269&    0.021& ---\\
   EC-129760&    267.62088&    -37.06300&    0.46336&         EC&   16.773&   16.405&   16.357&    0.023&    0.023&    0.024&    0.163&    0.168&    0.157&    0.042&    0.051&    0.048&    0.270&    0.022& ---\\
          C9&    267.52458&    -37.05575&    2.93268&        EC?&   15.329&   14.787&   14.547&    0.023&    0.023&    0.023&    0.213&    0.220&    0.168&    0.024&    0.030&    0.035&    0.210&    0.026& ---\\
         C16&    267.54762&    -37.01297&    0.24261&        EC?&   16.613&   16.285&   16.105&    0.024&    0.024&    0.025&    0.234&    0.229&    0.214&    0.068&    0.079&    0.100&    0.236&    0.012& ---\\
         C20&    267.52629&    -37.01319&    6.16530&        EC?&   15.038&   14.357&   14.195&    0.023&    0.022&    0.023&    0.147&    0.133&    0.140&    0.023&    0.022&    0.025&    0.236&    0.022& ---\\
         C27&    267.49183&    -37.01831&    2.33791&        EC?&   15.390&   14.724&   14.532&    0.027&    0.026&    0.024&    0.259&    0.254&    0.214&    0.037&    0.041&    0.034&    0.229&    0.021& ---\\
         C31&    267.51050&    -37.10583&   43.52855&        EC?&   13.747&   13.025&   12.981&    0.021&    0.022&    0.022&    0.110&    0.096&    0.095&    0.022&    0.022&    0.034&    0.250&    0.013& ---\\
\hline
\end{tabular}}
\end{center}
\footnotesize{{\bf Notes:} Star ID, coordinates (epoch J2000), period, variable type, and the pulsation mode are adopted from the literature \citep{clement2001, soszynski2014, alonsogarcia2021}. Type -- RRab: fundamental-mode RRL, RRc: overtone-mode RRL, T2C: Type II Cepheids, EC: Eclipsing binary, EC? - Eclipsing binary candidates. The columns 6-17 represent intensity-averaged mean magnitudes and their errors and peak-to-peak amplitudes and their uncertainties in the $JHK_s$ bands, respectively. The columns 18 and 19 display reddening values and their errors taken from the reddening maps of \citet{surot2020}. The last column shows alternate ID of the variable which was taken from the OGLE catalog of variable stars or the aforementioned references.}
\end{sidewaystable*}

\subsection{Proper motions and cluster membership of variables}

\begin{figure*}
\centering
\includegraphics[width=0.93\textwidth]{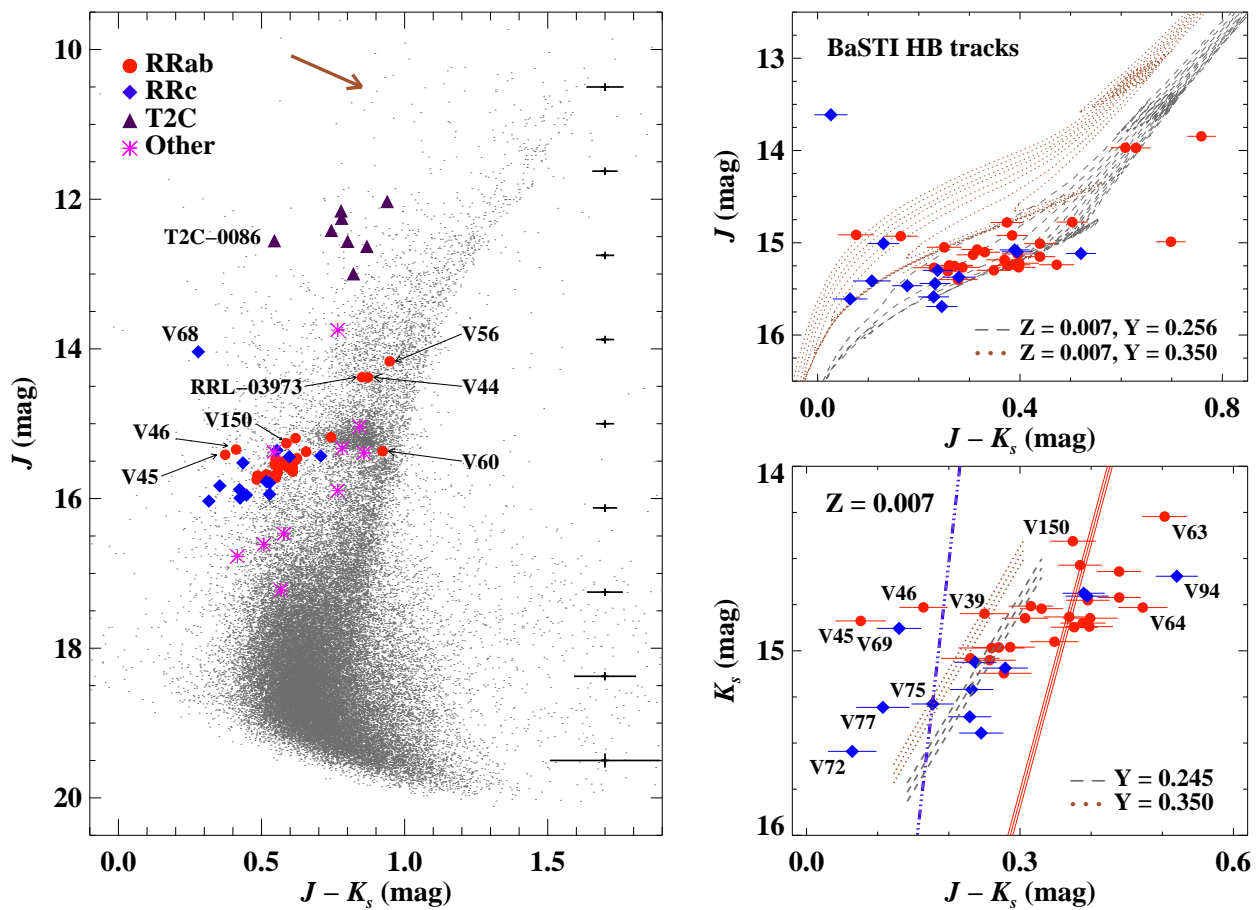}
\caption{ \textit{Left}: Variables in the NIR color--magnitude diagram of NGC 6441. The color--magnitude diagram is not corrected for extinction and the reddening vector is shown at the top of the panel. The representative error bars are $\pm5\sigma$ in magnitude and color. \textit{Top right}: Close-up of the RRL stars on the horizontal branch of the extinction-corrected ($J-K_s$), $J$ color--magnitude diagram. The horizontal branch stellar evolutionary models \citep{pietrinferni2006} for canonical helium (Y = 0.256) and enhanced helium content (Y = 0.350) are also shown. A distance modulus of 15.52~mag \citep{baumgardt2021} is adopted. \textit{Bottom right:} Close-up of the RRL stars in extinction-corrected ($J-K_s$), $K_s$ color--magnitude diagram. The dotted and dashed lines (with $\pm 1 \sigma$ errors due to distance and reddening uncertainties) represent predicted magnitudes and colors for RRL stars in our sample using theoretical PLZ relations of \citet{marconi2018} and  \citet{marconi2015} with and without helium term, respectively. 
The dashed-dotted blue and solid red lines represent the predicted blue edge of the first-overtone mode and the red edge of the fundamental mode, respectively. } 
\label{fig:cmd_all}
\end{figure*}

Our NIR catalogue was cross-matched with the {\it Gaia} data release 3 \citep[DR3,][]{prusti2016, vallenari2022} which resulted in 25987 common stars within $1\arcsec$ search radius. The mean proper motions along with the right ascension and declination axes are $\mu_\alpha = -2.56\pm0.02$ and $\mu_\delta = -5.31\pm0.02$ mas yr$^{-1}$. These agree well with the measurements based on the VVV proper motions \citep{alonsogarcia2021}. 
Fig.~\ref{fig:pms} displays the proper motions of all common stars with {\it Gaia}. Out of 60 variables, 48 have proper motions in {\it Gaia} DR3 which are also shown in Fig.~\ref{fig:pms}. Most variables fall within $\pm3\sigma$ of their mean proper motion values, where $\sigma\sim1.65~$mas yr$^{-1}$ is the standard deviation in the mean values along both axes. V60 and V154 have proper motion outside the 3 sigma scatter from the mean proper motion of the cluster. However, both these variables have a large re-normalized unit weight error (ruwe$~>5$), implying problems with their astrometric solutions. Given that majority of variables have unreliable astrometric measurements in the crowded cluster (27 out of 48 variables have ruwe$~>5$), we do not exclude any variable as a member of the cluster only based on their proper motions. \citet{alonsogarcia2021} used updated NIR proper motion and parallax catalogue of the VVV sources \citep{smith2018} for membership of variables in NGC 6441. Among 22 variables in common with our sample, \citet{alonsogarcia2021} classified V45, V68, V16, and V31 as field variables. 

\subsection{Variable stars on color--magnitude diagram}

Fig.~\ref{fig:cmd_all} shows the NIR color--magnitude diagram for NGC 6441. More than 50,000 point-like sources which were found in at least 5 science frames are plotted in the left panel. The color--magnitude diagram in the left panel is not corrected for extinction and some of the evident scatter could be due to differential reddening and field contamination. The red horizontal branch is well-populated and the bright T2C are clearly separated from fainter variables. Several RRL variables (V44, V56, V60, V68, RRL-03973) appear as outlier being either brighter or redder than the horizontal branch stars. \citet{layden1999} suggested that V44 (and V41) are probable RRL cluster members which appear brighter and redder due to photometric contamination. Unlike V41, V44 is significantly brighter in our photometry and is located within $3r_h$ of the cluster. Further inspection confirms that all outlier RRab (V44, V56, V60, and RRL-03973) are blended with a nearby bright source in the images. V60 also has a very small amplitude $(<0.1~\rm{mag})$ in all three $JHK_s$ bands. On the other hand, V68 is located well outside $7r_h$, and is a field RRc star as suspected by \citet{layden1999} and also classified by \citet{alonsogarcia2021}. Other variable candidates than RRL and T2C are either fainter or redder than the horizontal branch, except for  EC-128052 and C31. While EC-128052 is on the horizontal branch, it has a longer period (3.40209 days) than RRL. The bright unclassified variable, C31, is located in the outskirt, similar to V68, and is classified as field star \citep{alonsogarcia2021}. The longest-period ($P=1.0686$~days) RRab, V150, is also located at the bright end of the RRL horizontal branch. Note that \citet{corwin2006} derived a period of 0.529 days for V150, but phased NIR light curves exhibit large scatter with this period. While the period greater than 1 day suggests it maybe a short-period T2C, its location on color--magnitude diagram indicates it is still in the core helium burning evolutionary phase.

To correct the magnitudes and colors of variables for extinction, we obtained individual reddening values for each source based on the bulge reddening map of \citet{surot2020}{\footnote{\url{http://basti-iac.oa-teramo.inaf.it/vvvexmap/}}}. The reddening maps are based on NIR photometry of red clumps from the VVV survey. The resolution of this reddening map is $\sim1\arcmin$ in the region around NGC 6441. The estimated reddening values cover a range of $E(J-K_s)$ from $0.146$~mag to $0.312$~mag as shown in Table~\ref{tbl:var}. Adopting standard reddening law from \citet{card1989} and assuming $R_V\sim3.1$, \citet{gonzalez2012} provided estimates of extinction in NIR filters: $A_{J/H/K_s}=1.692/1.054/0.689 E(J-K_s)$. From \citet{nataf2013}, adopting $E(B-V)=E(V-I)/1.179$ and $E(V-I)=E(J-K_s)/0.407$, the mean value of $E(J-K_s)=0.235\pm0.038$~mag (or $E(B-V)=0.49$~mag) is consistent with the reddening value of $E(B-V)=0.47$~mag, as listed in \citet{harris2010}. The mean value of $E(J-K_s)$ also agrees well with the measurement of $0.18\pm0.06$ mag by \citet{alonsogarcia2021}.

The extinction corrected color--magnitude diagrams for RRL are shown in the right panels of Fig.~\ref{fig:cmd_all}.  In the top right panel, the $(J-K_s),~J$ color--magnitude diagram is shown together with the horizontal branch stellar evolutionary models from \citet{pietrinferni2006}. The canonical horizontal branch models with $\alpha$ enhancement ($[\alpha/\rm{Fe}]=0.4$~dex) from the BaSTI (a Bag of Stellar Tracks and  Isochrones){\footnote{\url{http://albione.oa-teramo.inaf.it/main_mod.php/}}} database are shown for Z = 0.007, $M=0.55-0.65M_\odot$, and a helium-to-metal enrichment ratio of $\Delta Y/\Delta Z=1.4$ (Y=0.256). Following the predictions of strong helium enrichment in NGC 6441 \citep[e.g.][]{busso2007, tailo2017}, we also adopted horizontal branch models corresponding to Y = 0.350. The canonical helium models within a narrow mass-range from $0.55M_\odot$ to $0.65M_\odot$ appear to populate the observed RRL color--magnitude distribution. However, the helium-enhanced evolutionary models for the same metallicity and mass-range are systematically brighter than the observed NIR magnitudes. Note that an increase in helium results in an increase in the luminosity level and, in turn, in the predicted pulsation periods of RRL stars.

In the bottom right panel of Fig.~\ref{fig:cmd_all}, the color--magnitude diagram is displayed in $(J-K_s),~K_s$ after excluding the obvious outliers in the left panel. Theoretically predicted boundaries of the RRL instability strip from \citet{marconi2015} are also shown after correcting for a distance modulus of $15.52$~mag \citep{baumgardt2021}. The brightest RRab (V63, $P=0.6978$~day) is possibly blended because its optical magnitude ($I=15.75$~mag) from OGLE is also significantly brighter than other similar period RRab stars (e.g., V62 with $P=0.6800$~day and $I=16.54$~mag). However, the brightest RRc (V94) is well-resolved in the images and it is brighter and redder than other RRc stars. \citet{pritzl2001} suspected that V94 is possibly a contact binary, but its classification as an RRc star is also supported by the OGLE light curves for this source (OGLE-BLG-RRLYR-03940). The magnitudes and colors of V94 indicate that it is a field variable. The long-period V150 ($P = $1.0686 days) is located within the instability strip. While a few RRL that fall outside the predicted boundaries may be outlier due to photometric uncertainties, the brighter end of RRL appear to be systematically redder than the predicted red edge. This is in contrast with more metal-poor clusters M53 and M15 \citep{bhardwaj2021,bhardwaj2021a} where the predicted boundaries cover all cluster RRL. Note that the predicted instability strip from \citet{marconi2015} is independent of metallicity in NIR. It is possible that there is a metallicity dependence in the NIR, but the reddening uncertainties may also cause these RRL to appear redder than the predicted boundaries. 

The predicted mean magnitudes and colors for the period range of RRL in our sample are also shown in the bottom right panel of Fig.~\ref{fig:cmd_all} for the representative mean metallicity ([Fe/H]$=-0.45$~dex or Z=0.007) of NGC 6441. The magnitudes and colors were estimated using the theoretical PLZ relations of \citet{marconi2015} for the primordial helium content (Y=0.245) and using PLZY relations of \citet{marconi2018} for the enhanced helium (Y=0.350). The predicted color--magnitude lines are also offset using the previously adopted distance modulus to NGC 6441. The prediction of PLZ relations for canonical helium content results in magnitudes which are on average brighter at a given color, and the colors that are  systematically bluer than the NIR observations of RRL stars. The predicted colors and magnitudes of helium enhanced RRL stars based on PLZY relations are bluer and brighter than the predictions from the models with canonical helium content. The observations show a larger discrepancy with respect to helium enhanced models. Among the RRL that are bluer than the predicted blue edge, only V46 ($P = $0.90449 days) and V69 ($P = $0.56119 days) have unusually long periods for RRab and RRc, respectively. V45 is the shortest period RRab in our sample, and its classification is further confirmed based on the shape of the OGLE light curves (OGLE-BLG-RRLYR-03886). \citet{alonsogarcia2021} found it to be a field variable which explains its bluer color and shorter period in comparison to RRab variables in NGC 6441 variables. Only V39 ($P = $0.82979 days) falls perfectly on the predicted magnitude and color strip using helium-enhanced models. The difference between empirical and predicted colors can partly be attributed to the uncertainties in adopted color-temperature transformations in models, but a larger reddening value will also result in redder and brighter magnitudes for RRL stars, thus, providing a better consistency with the theoretical predictions.   

\subsection{Period-amplitude diagrams}

\begin{figure}
\includegraphics[width=\columnwidth]{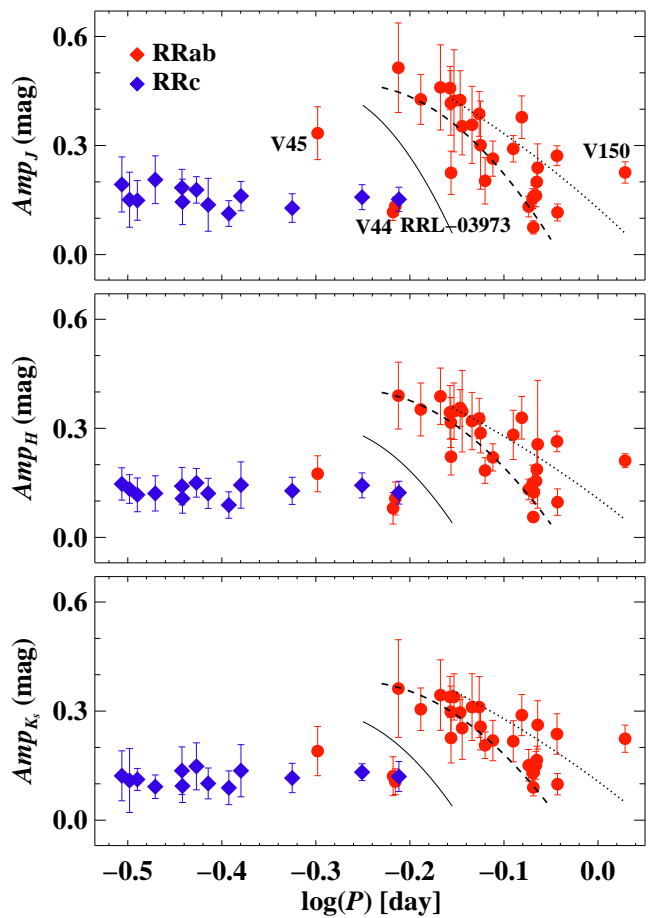}
\caption{NIR Bailey diagrams for RRL stars in NGC 6441. The solid and dashed lines correspond to the loci of OoI and OoII RRab from \citep{bhardwaj2020, bhardwaj2021}. The dotted line shows the locus of scaled $V$-band amplitudes of long-period OoIII type RRab taken from \citet{braga2020}.}
\label{fig:bailey_rrl}
\end{figure}

\begin{figure}
\includegraphics[width=\columnwidth]{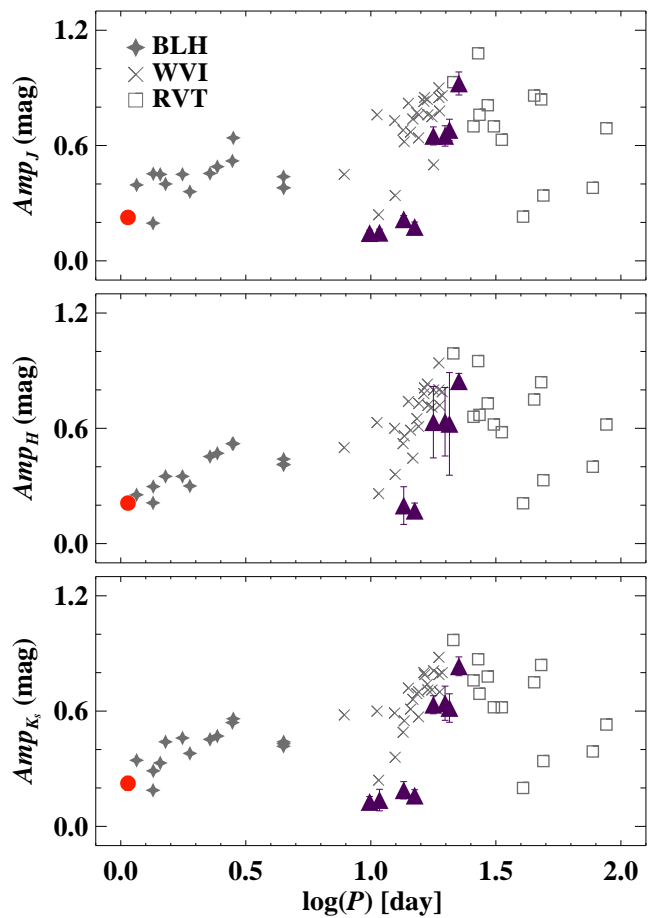}
\caption{NIR Period-amplitude diagrams for BL Herculis (BLH), W Virginis (WVI), RV Tauri (RVT) in Galactic GCs. The red circle represent V150 and triangles show T2Cs in NGC 6441.}
\label{fig:bailey_t2c}
\end{figure}

The period--amplitude or Bailey diagrams \citep{bailey1902} for RRL and T2Cs are useful for their separation into different subtypes and to investigate RRL in different Oosterhoff type GCs \citep{catelan2009}. Fig.~\ref{fig:bailey_rrl} displays NIR amplitudes of RRL variables in NGC 6441. The amplitudes are well determined and the error bars represent the scatter around the best-fitting templates which were used to determine peak-to-peak amplitudes. The monotonic decrease in the amplitudes of RRab as a function of period is evident except for the longest-period RRab (V150). The short-period RRab (V45), which is classified as a field variable in \citep{pritzl2003} and \citet{alonsogarcia2021}, is also an outlier in the period-amplitude diagram. Another two shorter-period RRab (V44 and RRL-03973) are likely blended and have small amplitudes for their periods, and were also identified as outliers in the color--magnitude diagrams. 

 It is known that RRL in NGC 6441 fall on the locus of metal-poor OoII type GCs in the optical Bailey diagrams \citep{pritzl2003}. As expected, no RRab appears to be close to OoI locus but occupy locations closer to OoII locus in NIR period-amplitude diagrams. However, the majority of long-period RRab fall between OoII or OoIII loci and no clear separation is evident in the Bailey diagrams. Apart from V150, only V46 and V56 have periods just above 0.9 day. But V56 is blended and has NIR amplitudes more than two times smaller than those of V46, which is closer to the locus of OoIII RRab. The NIR light curves and amplitudes of V150 are similar to a T2C (V92, $P=1.3$~days) in $\omega$ Cen which was proposed to be a candidate RRL \citep{braga2020}. However, no good quality optical light curves of V150 are available for its proper classification. The long periods of RRab could be due to the helium-enhancement which causes a systematic increase in the periods of RRL primarily due to increased luminosity levels for similar masses \citep{marconi2018a}. While the distribution of long-period RRL in the period-amplitude plane indeed suggests possible helium enhancement, the predicted magnitude and colors of helium enhanced RRL are brighter and bluer in the color-magnitude diagrams. 

Figure~\ref{fig:bailey_t2c} shows the period--amplitude diagrams for T2Cs in Galactic GCs with different mean metallicities \citep[see,][]{braga2020, bhardwaj2022}. The NGC 6441 variables are also shown including the long-period RRab, V150. The NIR amplitudes of V150 appear to be more consistent with those of BL Herculis variables than those of RRL stars shown in Fig.~\ref{fig:bailey_rrl}. This suggests that V150 may be a BL Herculis star in agreement with the period based classifications of T2Cs. However, its magnitudes and colors are more consistent with those of RRL stars. Out of 8 T2Cs in NGC 6441, two variables have periods between 20 and 23 days, and they are classified as RV Tauri variables \citep{soszynski2014}. The amplitudes of the longest-period T2C (V6) are consistent with those of known RV Tauri stars in GCs. But the NIR amplitudes of V126 ($P=20.61$~days) and other T2Cs with $P<20$~days agree well with those of known W Virginis variables in GCs. Although the amplitudes of variables with periods in the vicinity of 10 days are typically small, it is possible that the amplitudes of the centrally located T2Cs are also reduced due to crowding effects.  

\section{Period--luminosity relations} 
\label{sec:plrs}

\begin{table}
\begin{center}
\caption{NIR PLRs and PWRs of RRL in NGC 6441. \label{tbl:plrs}}
\begin{tabular}{cccccc}
\hline\hline
{Band} & {Type} & {$a_\lambda$} & {$b_\lambda$} & {$\sigma$}& {$N$}\\
\hline
     $J$ &  RRab &    14.960$\pm$0.029      &     $-1.987\pm$0.252      &      0.065 &   22\\
     $J$ &   RRc &    14.540$\pm$0.087      &     $-2.270\pm$0.228      &      0.076 &    9\\
     $J$ &   All &    14.982$\pm$0.019      &     $-1.777\pm$0.100      &      0.054 &   27\\
     $H$ &  RRab &    14.687$\pm$0.026      &     $-2.171\pm$0.221      &      0.059 &   21\\
     $H$ &   RRc &    14.236$\pm$0.083      &     $-2.629\pm$0.212      &      0.068 &    9\\
     $H$ &   All &    14.672$\pm$0.020      &     $-2.290\pm$0.107      &      0.060 &   29\\
   $K_s$ &  RRab &    14.609$\pm$0.021      &     $-2.418\pm$0.184      &      0.046 &   18\\
   $K_s$ &   RRc &    14.139$\pm$0.059      &     $-2.770\pm$0.152      &      0.050 &    9\\
   $K_s$ &   All &    14.604$\pm$0.014      &     $-2.438\pm$0.074      &      0.040 &   25\\
   \hline
      $W_{J,H}$ &  RRab &    14.188$\pm$0.032      &     $-2.709\pm$0.284      &      0.078 &   22\\
     $W_{J,H}$ &   RRc &    13.713$\pm$0.110      &     $-3.257\pm$0.289      &      0.092 &    9\\
    $W_{J,H}$ &   All &    14.158$\pm$0.023      &     $-3.130\pm$0.132      &      0.072 &   29\\
     $W_{H,K_s}$ &  RRab &    14.459$\pm$0.038      &     $-2.567\pm$0.327      &      0.079 &   15\\
     $W_{H,K_s}$ &   RRc &    13.722$\pm$0.186      &     $-3.567\pm$0.464      &      0.133 &    8\\
     $W_{H,K_s}$ &   All &    14.457$\pm$0.033      &     $-2.713\pm$0.166      &      0.087 &   22\\
   $W_{J,K_s}$ &  RRab &    14.327$\pm$0.034      &     $-2.886\pm$0.292      &      0.070 &   17\\
   $W_{J,K_s}$ &   RRc &    13.866$\pm$0.119      &     $-3.108\pm$0.307      &      0.099 &    9\\
   $W_{J,K_s}$ &   All &    14.317$\pm$0.028      &     $-2.922\pm$0.147      &      0.079 &   26\\
\hline
\end{tabular}
\end{center}
\footnotesize{{\bf Notes:} The zero-point ($a$), slope ($b$), dispersion ($\sigma$) and 
the number of stars ($N$) in the final PLR fits are tabulated.}
\end{table}

\begin{figure*}
\centering
\includegraphics[width=0.97\textwidth]{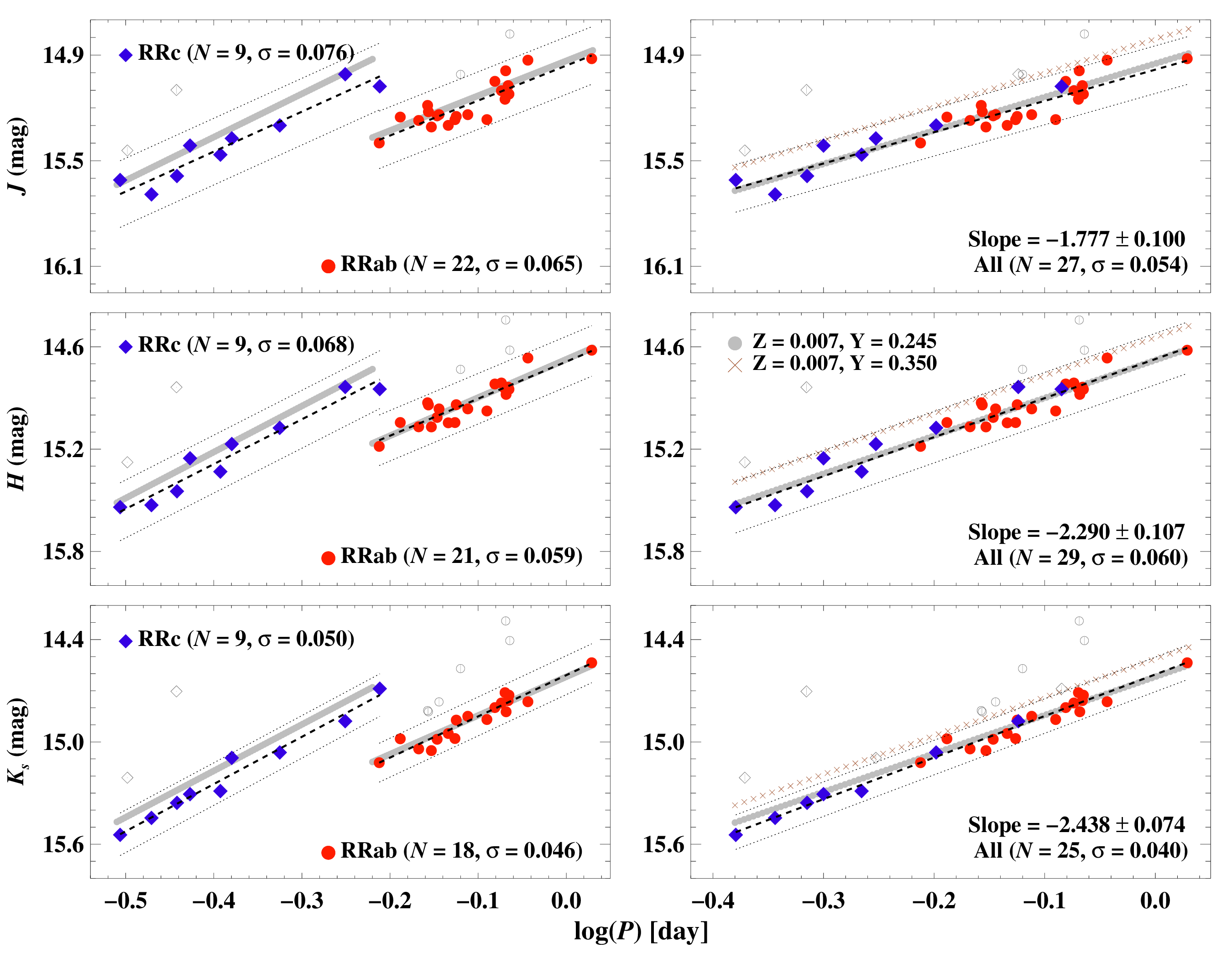}
\caption{NIR period-luminosity relations for RRab and RRc stars (left) and all RRL stars (right) in $J$ (top), $H$ (middle), and $K_s$ (bottom). In the right panels, the periods for the RRc stars have been shifted to their corresponding fundamental-mode periods, as explained in the text. The dashed lines represent best-fitting linear regressions over the period range under consideration while the dotted lines display $\pm 2.5\sigma$ offsets from the best-fitting PLRs. The shaded (grey) line and crossed (brown) lines represent theoretically predicted PLRs for metallicities representative of NGC 6441 but with canonical and enhanced helium, respectively.} 
\label{fig:rrl_plr}
\end{figure*}

\begin{figure*}
\begin{center}
\includegraphics[width=0.95\textwidth]{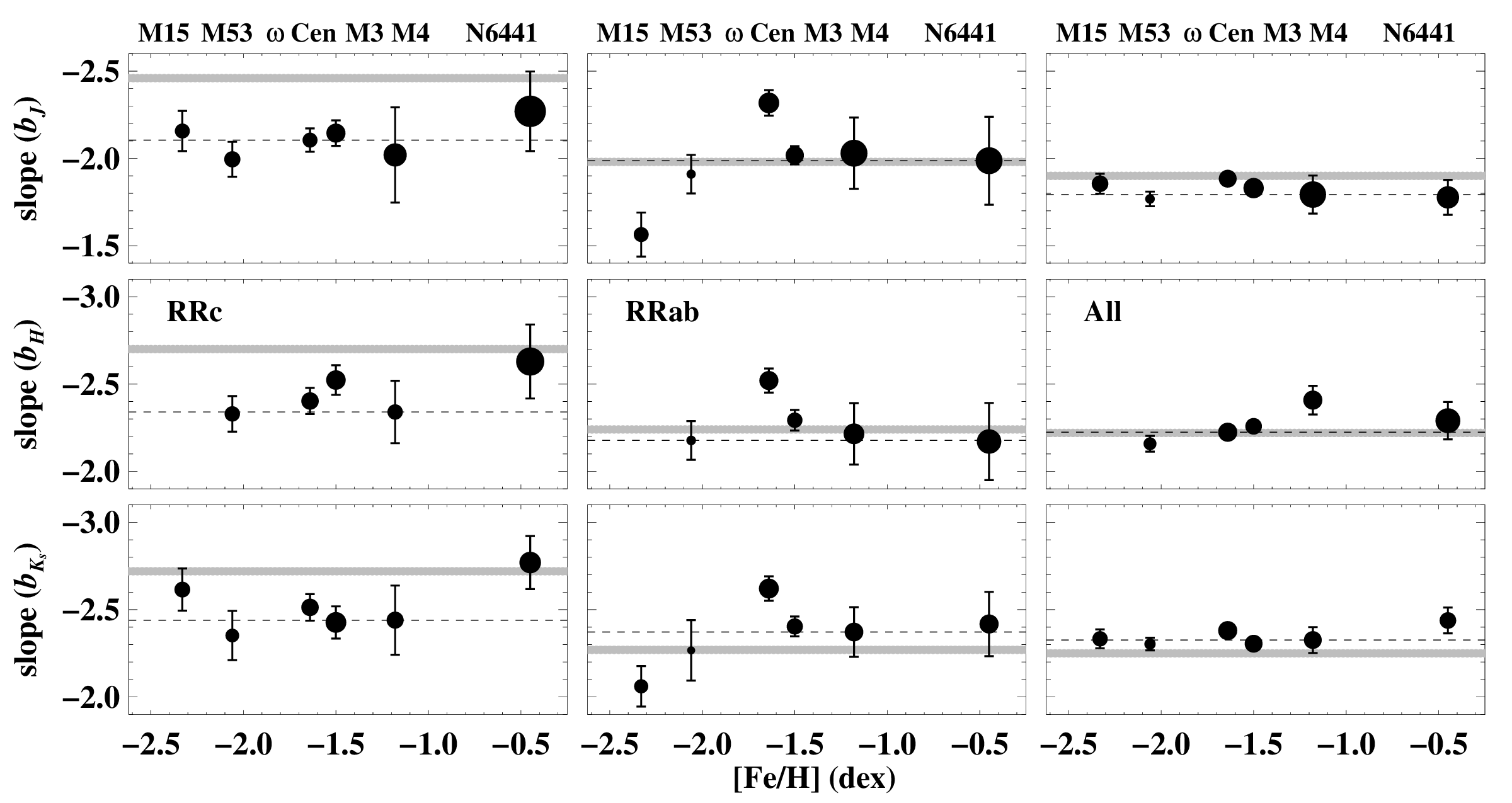}
\caption{The slopes of $JHK_s$ PLRs for RRc, RRab, and global sample of all RRL in GCs of different mean metallicities. The slopes are taken from the following studies: M15 \citep[{[Fe/H]}$\sim-2.33$~dex,][]{bhardwaj2021a},  M53 \citep[{[Fe/H]}$\sim-2.06$~dex,][]{bhardwaj2021}, $\omega$ Cen \citep[{[Fe/H]}$\sim-1.64$~dex,][]{braga2018}, M3 \citep[{[Fe/H]}$\sim-1.50$~dex,][]{bhardwaj2020}, M4 \citep[{[Fe/H]}$\sim-1.18$~dex,][]{braga2015}, and NGC 6441 ([Fe/H]$\sim-0.45$~dex, this work). The increase in symbol size represents an increase in the dispersion in the PLRs. The shaded (grey) lines represent the slopes of the theoretically predicted PLZ relations from \citet{marconi2015}. The dashed lines represent the median value of slopes in each panel.} 
\label{fig:rrl_slopes}
\end{center}
\end{figure*}

\subsection{RRL Period--Luminosity and Period--Wesenheit relations}

RRL stars are known to follow tight PLRs in NIR with a strong dependence on metallicity \citep{catelan2004, marconi2015, bhardwaj2021}. RRL stars in metal-rich NGC 6441 offer a test for the universality of PLRs when compared with the PLRs in more metal-poor GCs. Before deriving the PLRs, we exclude 7 variables (V44, V45, V56, V63, V68, V94, and RRL-03973) which have been discussed as outliers in either the pulsation properties or kinematics of cluster RRL. The extinction-corrected mean magnitudes from our template-fitted light curves were used to derive PLRs of the following form:
\begin{equation}
{m_\lambda} = a_\lambda + b_\lambda\log(P),
\label{eq:plr}
\end{equation}

\noindent where $b_\lambda$ and $a_\lambda$ are the slope and zero-point of the best-fitting PLR for a given wavelength. The individual samples of RRab and RRc as well as the combined sample of all RRL were used to derive PLRs. For the global sample of RRL, pulsation periods of first-overtone mode RRc stars were fundamentalized using the equation: $\log(P_{\textrm{RRab}})=\log(P_{\textrm{RRc}})+0.127$ \citep{petersen1991, coppola2015}. Our final sample to derive PLRs consists of 35 RRL variables including 24 RRab and 11 RRc stars.

Figure~\ref{fig:rrl_plr} displays $JHK_s$ PLRs for RRab, RRc, and all RRL stars in our final sample. We iteratively removed the single largest outlier from an initial linear regression until all residuals were within $2.5\sigma$ around the best-fitting relation. Table~\ref{tbl:plrs} lists the results of the best-fitting linear regression. Despite the small number of RRL in NGC 6441, the slope and zero-points are well constrained. In the case of RRc stars, V106 exhibits the largest residuals in $JHK_s$ PLRs. V106 is centrally located and is possibly bright due to blending with a nearby bright source. For the same reason, several RRab stars were also found to be brighter than the best-fitting PLRs. The results listed in Table~\ref{tbl:plrs} show that the slopes of the PLRs are the steepest for RRc stars. We also exclude the long-period V150 from the analysis and found that the slopes and zero-points are statistically consistent. The difference in the zero-points of PLRs with and without V150 is smaller than $0.011$ mag for the global sample of RRL in $JHK_s$ bands. This further strengthens the classification of V150 as RRab despite it having a period longer than 1 day. The relatively larger scatter in the RRL PLRs than those in other GCs (e.g. M3, M53, M15) can be attributed to uncertainties due to differential reddening, the photometric uncertainties, and a possible metallicity spread \citep[][]{clementini2005} in NGC 6441. However, we also note that \citet{gratton2007} did not find any significant metallicity spread in NGC 6441. 

Theoretically predicted NIR PLRs for different samples of RRL are also shown as shaded regions in Fig.~\ref{fig:rrl_plr}. The absolute magnitudes were shifted with a distance modulus of $15.52$~mag \citep{baumgardt2021} to NGC 6441. The predicted relations are from \citet{marconi2015} for the mean metallicity representative of NGC 6441. These models computed with a canonical helium content are in good agreement with observations except for the $JHK_s$ band PLRs for RRc stars. In the right panels, theoretical PLRs from \citet{marconi2018a} computed including the enhanced helium are also shown. The PLRs with enhanced helium (Y = 0.350) are systematically brighter than the observed PLRs which further suggests that RRL stars in NGC 6441 may not have such high helium content. The RRL which were also located outside the instability strip in the color--magnitude diagram shown in the bottom right panel of Fig.~\ref{fig:cmd_all}, exhibit residuals within $2.5\sigma$ in all three $JHK_s$ PLRs. Among long-period RRL stars, only V60 and V143 are outliers in the PLRs. While V60 is likely blended (see Fig.~\ref{fig:cmd_all}), V143 is located on the red edge of the instability strip. 

We also derived reddening-free Wesenheit magnitudes in NIR bands to investigate the impact of reddening. Following \citet{marconi2015}, we adopted the \citet{card1989} reddening law and derived Wesenheit magnitudes: $W_{J,H}=H-1.68(J-H)$, $W_{H, K_s}=K_s-1.87(H-K_s)$, and $W_{J,K_s}=K_s-0.69(J-K_s)$, for our sample of RRL stars. We derived PWRs in NIR bands in the form of equation~(\ref{eq:plr}), and the results of the  best-fitting linear regressions are tabulated in Table~\ref{tbl:plrs}. It is evident from Table~\ref{tbl:plrs} that the scatter in the PWRs increases as compared to $JHK_s$ PLRs. The scatter in $W_{H,K_s}$ Wesenheit relations, in particular for RRc stars, is quite large presumably due to the narrow color range in ($H-K_s$). The relatively narrow range of NIR colors of RRL and their larger uncertainties together with possible variation in the reddening law towards the cluster are likely contributing to the increase in the scatter in the PWRs of RRL stars. If we assume a different reddening law of \citet{nishiyama2009}, empirically determined zero-points of NIR PLRs are fainter than those of the predicted theoretical relations based on models with or without helium enhancement, but no significant variation is seen in the slopes of the PLRs.

Figure~\ref{fig:rrl_slopes} displays the slopes of the PLRs for the RRc, RRab, and the global sample of RRL stars in GCs of different mean metallicities. Only GCs with time-series NIR photometry in at least two filters are shown. For all samples of RRL, the slopes are best-constrained in the clusters with the largest number of variables (e.g., M3 and $\omega$ Cen). The slopes of RRc stars are typically the steepest and are also consistent between different GCs. Interestingly, the slopes of RRab stars in the most metal-poor ([Fe/H]$<-2.0$~dex) GCs appear to be shallower than those in more metal-rich systems. At face value, the slopes of $JK_s$ band PLRs for RRab stars in the metal-poor GC M15 and the metal-rich NGC 6441 are different, but they are statistically consistent within their uncertainties. Nevertheless, more metal extreme GCs are needed to confirm this possible trend in the slope of RRab stars. The RRab stars in $\omega$ Cen show the steepest slope of the PLRs which becomes consistent with those in the metal-rich GCs when the  metallicity term is included \citep[see,][]{braga2018}. In comparison with the theoretical models, the average slopes of RRc stars are significantly shallower than the predicted slopes in $JHK_s$ filters. However, the slopes of the PLRs for the global sample of RRL are consistent in all GCs and with theoretically predicted PLZ relations for the entire range of metallicity. Therefore, the use of the global sample of RRL is recommended under the assumption of the universality of PLRs to avoid any biases in RRL-based distance measurements.

\subsection{Distance and reddening}

Multiband NIR photometry for RRL in NGC 6441 can be used to estimate distance and reddening to the cluster, simultaneously. The apparent distance modulus and the extinction in $K_s$-band are estimated by simultaneously applying the following equation for different combinations of two NIR bandpasses:
\begin{equation}
    \mu + A_\lambda = m_\lambda - M_\lambda, 
\label{eq:mu_ak}
\end{equation}

\noindent where, $\mu$ is the true distance modulus and $A_\lambda$, $m_\lambda$, and $M_\lambda$ correspond to the extinction, apparent magnitude, and absolute magnitude in a given $J/H/K_s$ filter. The intensity-averaged mean magnitudes for RRL in NGC 6441 are provided in Table~\ref{tbl:data} and the absolute magnitudes can be estimated using the calibrated PLZ relations.  The extinction ratios ($A_J/A_{K_s}=2.46$, $A_H/A_{K_s}=1.53$), which were derived from the total-to-selective absorption values mentioned in Section 3.1, were used to convert extinction in $J$ and $H$ bands to $A_{K_s}$. We used equation~(\ref{eq:mu_ak}) in any two NIR filters to solve for the  $K_s$-band extinction and distance modulus to NGC 6441.

\begin{figure}
\includegraphics[width=\columnwidth]{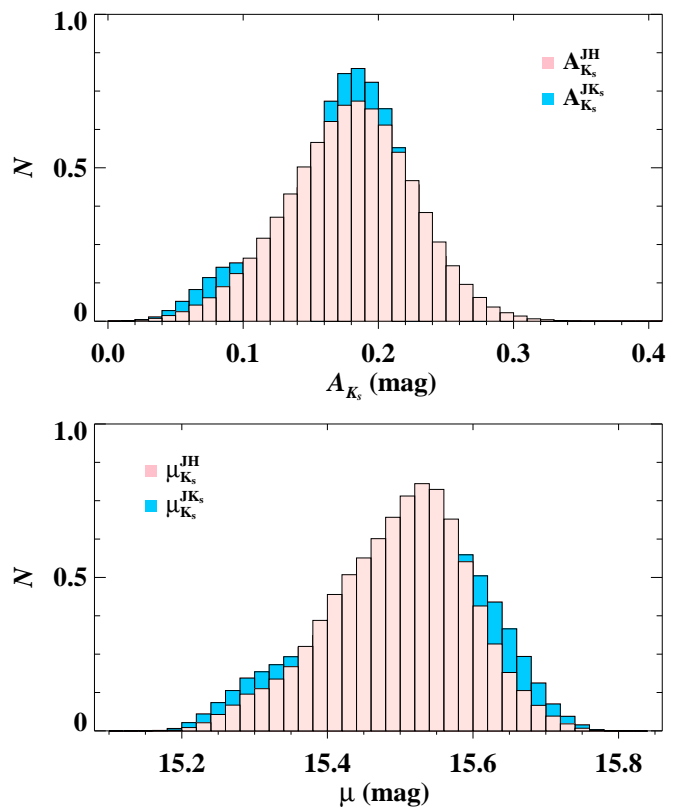}
\caption{Normalized histograms of extinction ($A_{K_s}$) and distance modulus ($\mu$), estimated simultaneously using PLZ relations in two NIR filters, are shown in the top and bottom panels, respectively.} 
\label{fig:mu_ak}
\end{figure}

\begin{table}
\begin{center}
\caption{Distance modulus and extinction to NGC 6441. \label{tbl:mu_ak}}
\begin{tabular}{ccc}
\hline\hline
            &   $\mu$ (mag)    &  $A_{K_s}$ (mag) \\
\hline
$J,H$    &   15.510$\pm$0.015$\pm$0.091    &  0.180$\pm$0.010$\pm$0.045    \\
$J,K_s$  &   15.516$\pm$0.017$\pm$0.109    &  0.180$\pm$0.009$\pm$0.042    \\
$H,K_s$  &   15.524$\pm$0.022$\pm$0.131    &  0.177$\pm$0.016$\pm$0.084   \\
$W_{JH}$    &   15.508$\pm$0.022$\pm$0.090  &   \\
$W_{HK_s}$  &   15.517$\pm$0.029$\pm$0.129  &   \\
$W_{JK_s}$  &   15.526$\pm$0.022$\pm$0.093  &   \\
\hline
\multicolumn{3}{c}{$A_{K_s}$ determined using $E(J-K_s)$ values from Table~\ref{tbl:var}}\\
\hline
$J$         &   15.534$\pm$0.018$\pm$0.066  &   \\
$H$         &   15.520$\pm$0.017$\pm$0.065  &   \\
$K_s$       &   15.533$\pm$0.015$\pm$0.052  &   \\
\hline
\end{tabular}
\end{center}
\footnotesize{{\bf Notes -} The two errors listed for each measurement represent the statistical uncertainty and the scatter in the mean values, respectively.}
\end{table}

Theoretical calibrations from \citet{marconi2015} were adopted to determine absolute magnitudes for each RRL variable. The global sample of RRL was used for this analysis since the slopes of their PLRs are consistent between theory and observations for the entire metallicity range as shown in Fig.~\ref{fig:rrl_slopes}. All outliers of the PLRs, as discussed in the previous subsection, were excluded from the analysis. We generated $10^4$ random realizations of our estimates by varying the apparent and absolute magnitudes within their uncertainties in each iteration. The errors on mean apparent magnitudes are listed in Table~\ref{tbl:data} which were supplemented by adding the extinction correction uncertainties in quadrature. The uncertainties on the absolute magnitude were estimated by including the errors in the zero-point of theoretical PLZ relations, error in the adopted mean-metallicity \citep[$\Delta\rm{[Fe/H]}=0.06$~dex,][]{clementini2005}, and the error on the slope with respect to the difference in the mean period of RRL in theoretical calibrations and in NGC 6441. 

Fig.~\ref{fig:mu_ak} displays the normalized histograms of $10^4$ determinations of extinction in $K_s$-band and the true distance modulus to the cluster for RRL in our sample. The estimates based on $JH$ and $JK_s$ filter combinations are well constrained. We do not consider distance and reddening estimates based on $H$ and $K_s$ filters for this analysis because the $W_{H,K_s}$ PWR exhibits a relatively large scatter. Therefore, a narrow range in ($H-K_s)$ color does not provide a strong constraint on the reddening values. Indeed, a larger scatter is evident in the mean values of reddening and distance based on $HK_s$ filters in Table~\ref{tbl:mu_ak}. Nevertheless, the mean values of the extinction and the distance modulus listed in Table~\ref{tbl:mu_ak} are in excellent agreement. We exclude the measurements based on $HK_s$ filters and obtain an average extinction in $K_s$-band, $A_{K_s}=0.180\pm0.043$~mag, and a distance modulus, $\mu=15.513\pm0.100$~mag, to NGC 6441. The quoted uncertainties represent the scatter in the mean values due to the errors in the photometry, reddening, adopted mean metallicity, and the theoretical calibrations. The mean $A_{K_s}$ value corresponds to $E(J-K_s)=0.261\pm0.062$~mag, and is statistically consistent with the reddening values determined by \citet[][$E(J-K_s)=0.207\pm0.009$~mag]{alonsogarcia2021}. We independently estimate the individual reddening values for all sources in NGC 6441 using the reddening maps of \citet{surot2020} based on red clump stars. We find an average reddening, $E(J-K_s)=0.245\pm0.036$~mag, in excellent agreement with our measurements based on RRL stars. 

\begin{figure}
\includegraphics[width=\columnwidth]{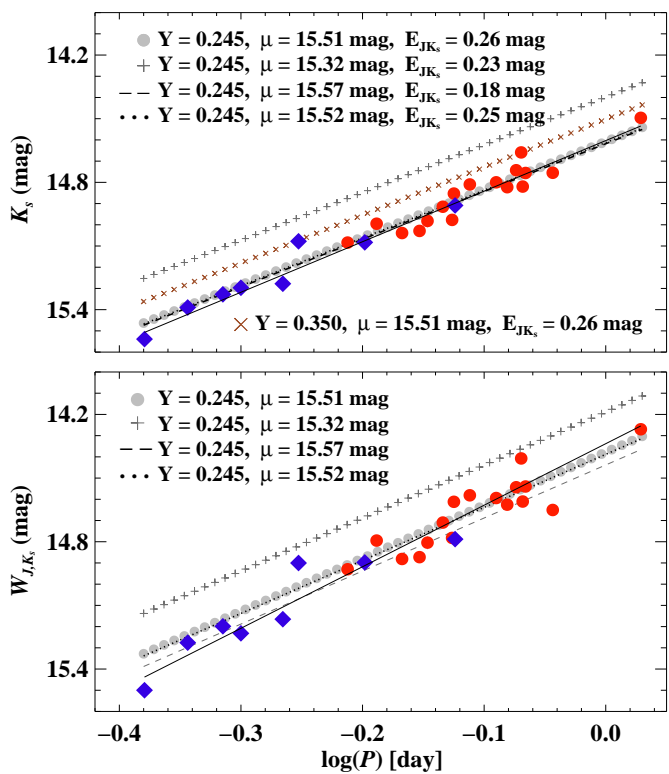}
\caption{Comparison of empirical $K_s$-band PLR (top) and $W_{J,K_s}$ PWR (bottom) for RRL stars with the theoretical predictions together with different sets of adopted composition, distance, and reddening values.} 
\label{fig:k_wjk}
\end{figure}

The Wesenheit relations can be used to obtain distance independent of reddening estimates for RRL in our sample. We used theoretical PWZ relations from \citet{marconi2015} to obtain a distance modulus to NGC 6441 and the results are listed in Table~\ref{tbl:mu_ak}. Interestingly enough, the distance moduli are in excellent agreement with those determined by solving equation~(\ref{eq:mu_ak}) in any two NIR filters. Therefore, we adopt an average reddening value of $E(J-K_s)=0.261\pm0.062$~mag and a distance modulus of  $\mu=15.513\pm0.016~\rm(stat.)\pm0.100~(syst.)$~mag to NGC 6441. Our RRL based distance modulus is relatively smaller than the previous determinations based on NIR photometry of RRL stars by \citet[][$15.67\pm0.07$~mag]{dallora2008} and \citet[][$15.57\pm0.05$~mag]{alonsogarcia2021}. We also used the individual reddening values determined from the reddening maps of \citet{surot2020} to derive a distance modulus to NGC 6441 using theoretical PLZ calibrations. The results listed in Table~\ref{tbl:mu_ak} also agree well with our adopted distance modulus within their quoted uncertainties. Our final distance of $12.67\pm0.09$~kpc to NGC 6441 is in excellent agreement with the latest recommended distance of $12.73\pm0.16$~kpc  in the catalog of \citet{baumgardt2021} based on {\it Gaia}, {\it HST} and literature data.

Fig.~\ref{fig:k_wjk} displays the comparison of empirical $K_s$-band PLR and $W_{J,K_s}$ PWR with theoretical predictions combined with different estimates of distance and reddening values. The predicted relations for canonical helium content agree well with observations when we use our estimated reddening and distance modulus. Similarly, the distance modulus and reddening estimate by \citet{alonsogarcia2021} result in an agreement between theoretical and empirical relations. However, the PLR and PWR based on enhanced helium RRL models are significantly brighter than the empirically derived NIR relations regardless of the adopted distance or reddening. Furthermore, a smaller distance from \citet{harris2010} catalog also provides significantly brighter predicted zero-points which do not agree with the observations. Our distance and reddening estimates are in excellent agreement with those adopted for color-magnitude diagrams in Fig.~\ref{fig:cmd_all}. Therefore, the consistency of NIR observations with canonical helium models on the color--magnitude and PLR planes suggests that the RRL in NGC 6441 are not significantly helium enhanced, and it is likely that the amount of helium enhancement is much smaller (Y $<<$ 0.35) than suggested in previous studies \citep[][]{busso2007, bellini2013, tailo2017}. However, it is also possible that the impact of the helium enrichment on NIR pulsation properties, in particular, on magnitudes and colors of RRL is significantly smaller than the predictions of the pulsation models. A multiwavelength comparison of observations of RRL in NGC 6441 with the evolutionary and pulsation models may provide a better constraint on the possible helium enrichment and an insight into the peculiar horizontal branch morphology in this cluster. 

\subsection{Type II Cepheid period-luminosity relations}

\begin{figure}
\includegraphics[width=\columnwidth]{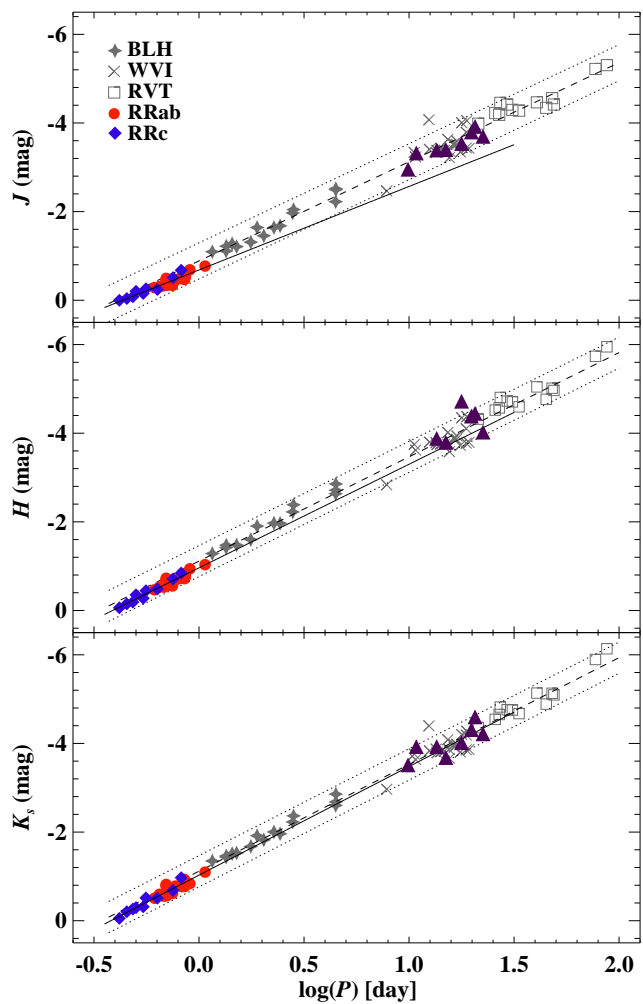}
\caption{NIR PLRs for T2Cs in Galactic GCs. T2Cs in NGC 6441 are shown as filled triangles. The dashed lines represent best-fitting linear regressions to T2C sample while the dotted lines display $\pm 2.0\sigma$ offsets from the best-fitting PLRs. The solid line is the best-fitting linear regression to the global sample of RRL stars but is extended to T2Cs in NGC 6441.} 
\label{fig:t2c_plr}
\end{figure}

Our sample of 8 T2Cs in NGC 6441 is the largest with NIR photometry followed by 7 T2Cs in $\omega$ Cen \citep{navarrete2017}. However, unlike $\omega$ Cen T2Cs which cover a wide period range, T2Cs in NGC 6441 are within a narrow period range ($0.99<\log(P)<1.35$~days). Therefore, we do not fit a PLRs to our sample of T2Cs. Nevertheless, T2Cs in metal-rich NGC 6441 provide an opportunity to investigate possible metallicity dependence on their PLRs when compared with variables in the metal-poor GCs. We use the updated T2C sample compiled in \citet{braga2020} to derive PLRs. The sample was further updated with NIR photometry for 2 T2Cs in NGC 7078 from \citet{bhardwaj2021a}. For NGC 6441 T2Cs, we adopted the distance and reddening values estimated using RRL in the previous subsection. 

Fig.~\ref{fig:t2c_plr} displays $JHK_s$ PLRs for T2Cs in GCs. All T2Cs in NGC 6441 are located within $2\sigma$ of the best-fitting PLRs in $J$ and $K_s$-bands. Only one variable is outside this limit in $H$-band. All these T2Cs are fainter than the best-fitting linear regression in $JHK_s$-bands if we adopt a smaller distance from \citet{harris2010} with our reddening estimates. Since the residuals do not exhibit any dependence on mean metallicity of the GCs, the metal-rich T2Cs are unlikely to be fainter than metal-poor T2Cs - as seen for the RRL stars. This consistency between T2Cs in NGC 6441 and other GCs in the PLRs further confirms the accuracy of our distance and reddening values based on RRL stars.

RRL and T2Cs are known to follow similar PLRs in NIR \citep{matsunaga2006, bhardwaj2017a, braga2020}. Since RRL PLRs exhibit a dependence on metallicity, we correct their absolute magnitudes for the mean-metallicity of NGC 6441 using the coefficient of theoretical PLZ relations. Fig.~\ref{fig:t2c_plr} also compares the best-fitting linear regressions to RRL and T2Cs. The slope of RRL PLR in $J$-band is shallower than that for T2Cs in GCs, but the $H$-band slopes are statistically consistent. In the $K_s$-band, the RRL and T2Cs perfectly follow the same PLR provided the metallicity term is accounted for RRL stars. The similarity of RRL and T2Cs PLRs in metal-rich NGC 6441 and in metal-poor $\omega$ Cen \citep{braga2020} strengthens the universality of combined $K_s$-band PLR of these population II distance indicators. 
Furthermore, the consistency between RRL and T2Cs PLRs in $K_s$-band also suggests that T2Cs PLRs do not show any significant dependence on metallicity. A more detailed analysis of metallicity dependence will be presented in a future study with more T2Cs in similar metal-rich bulge globular cluster NGC 6388. 

\section{Conclusions}
\label{sec:discuss}

We presented NIR time-series observations of variable stars in NGC 6441 and investigated pulsation properties of population II distance indicators, RRL and T2C variables. The multiband NIR photometry for RRL in NGC 6441 was used to provide new estimates of reddening and distance to the cluster. The theoretical calibrations of RRL PLZ relations were used to simultaneously estimate the mean reddening, $E(J-K_s)=0.261\pm0.062$~mag, and the distance, $d=12.67\pm0.09$~kpc, to NGC 6441 by accounting for the errors in photometry, reddening, metallicity, and the calibrator relations. Our reddening estimates are in excellent agreement with the independent value from the reddening maps of \citet{surot2020} based on red clumps and the measurement by \citet{alonsogarcia2021} using VVV photometry. Similarly, the RRL based distance agrees well with the latest measurement based on {\it Gaia} and literature data \citep{baumgardt2021}. Furthermore, we showed that our distance and reddening values lead to an excellent agreement between T2Cs in NGC 6441 and other GCs on the NIR PLRs. We also find that RRL and T2Cs follow the same PLR in the $K_s$-band which confirms the consistency of the combined PLRs of these population II distance indicators in stellar systems with different metallicities.

The NIR  pulsation properties of RRL in the metal-rich cluster NGC 6441 are consistent with theoretical predictions for the mean metallicity of the cluster and the adopted canonical helium content. In particular, the observed magnitudes and colors in NIR do not agree with the predictions of helium enhanced pulsations models. In the color--magnitude diagrams, the predicted magnitudes and colors based on helium enhanced models are respectively brighter and bluer than the canonical helium models and the observations. Similarly, the helium enriched models are systematically brighter in the NIR PLRs while the theoretical relations with canonical helium content agree well with the observations. While the unusually long periods of RRL in NGC 6441 are thought to be due to the helium enhancement, the impact of such enhancement, if present, is minimal in NIR as compared to the predictions of the pulsation models. On the contrary, if the predictions of pulsation models are correct, this would suggest that evolutionary effects or differential reddening at shorter wavelengths may be contributing more to the peculiar tilted horizontal branch morphology than previously thought. More detailed multiwavelength investigations of NGC 6441 are warranted to gain further insights into the horizontal branch and RRL populations of this intriguing bulge globular cluster.

\begin{acknowledgements}
     This project has received funding from the European Union’s Horizon 2020 research and innovation programme under the Marie Skłodowska-Curie grant agreement No. 886298. Support for M.C. is provided by the Ministry for the Economy, Development, and Tourism's Millennium Science Initiative through grant ICN12\textunderscore 12009, awarded to the Millennium Institute of Astrophysics (MAS), and by Proyecto Basal ACE210002 and FB210003.
\end{acknowledgements}

%
%
\bibliographystyle{aa}
\bibliography{mybib_final.bib}

\end{document}